\documentclass[twocolumn,trackchanges]{aastex701}

\newcommand\aastex{AAS\TeX}

\usepackage{amsmath}
\usepackage{multirow}
\usepackage{soul}
\shorttitle{Mass Transfer Stability of Massive Helium Binary Stars}
\shortauthors{Zhang et al.}
\graphicspath{{./}}

\begin{document}

\title{Template \aastex v7.0.1 Article with Examples\footnote{Footnotes can be added to titles}}

\title{Adiabatic Mass Loss In Binary Stars. VI. Massive Helium Binary Stars}

\author[orcid=0009-0001-3638-3133]{Lifu Zhang}
\affiliation{Yunnan Observatories, Chinese Academy of Sciences, Kunming 650216, People's Republic of China}
\affiliation{International Centre of Supernovae (ICESUN), Yunnan Key Laboratory of Supernova Research, Kunming 650216, P.R. China}
\affiliation{University of Chinese Academy of Sciences, Beijing 100049, People's Republic of China}
\email{zhanglifu@ynao.ac.cn}

\author[orcid=0000-0002-6398-0195]{Hongwei Ge}
\affiliation{Yunnan Observatories, Chinese Academy of Sciences, Kunming 650216, People's Republic of China}
\affiliation{International Centre of Supernovae (ICESUN), Yunnan Key Laboratory of Supernova Research, Kunming 650216, P.R. China}
\affiliation{University of Chinese Academy of Sciences, Beijing 100049, People's Republic of China}
\email[show]{gehw@ynao.ac.cn}

\author[orcid=0000-0002-1421-4427]{Zhenwei Li}
\affiliation{Yunnan Observatories, Chinese Academy of Sciences, Kunming 650216, People's Republic of China}
\affiliation{International Centre of Supernovae (ICESUN), Yunnan Key Laboratory of Supernova Research, Kunming 650216, P.R. China}
\affiliation{University of Chinese Academy of Sciences, Beijing 100049, People's Republic of China}
\email{lizw@ynao.ac.cn}

\author[orcid=0009-0006-9211-2860]{Hailiang Chen}
\affiliation{Yunnan Observatories, Chinese Academy of Sciences, Kunming 650216, People's Republic of China}
\affiliation{International Centre of Supernovae (ICESUN), Yunnan Key Laboratory of Supernova Research, Kunming 650216, P.R. China}
\affiliation{University of Chinese Academy of Sciences, Beijing 100049, People's Republic of China}
\email{chenhl@ynao.ac.cn}

\author[orcid=0000-0003-4265-7783]{Dengkai Jiang}
\affiliation{Yunnan Observatories, Chinese Academy of Sciences, Kunming 650216, People's Republic of China}
\affiliation{International Centre of Supernovae (ICESUN), Yunnan Key Laboratory of Supernova Research, Kunming 650216, P.R. China}
\affiliation{University of Chinese Academy of Sciences, Beijing 100049, People's Republic of China}
\email{dengkai@ynao.ac.cn}

\author[0000-0002-3839-4864]{Guoliang Lü}
\affiliation{School of Physical Science and Technology, Xinjiang University, Urumqi 830046, China}
\email{guolianglv@sina.com}

\author[0000-0002-7334-2357]{Xiaofeng Wang}
\affiliation{Physics Department, Tsinghua University, Beijing, 100084, China}
\email{wang_xf@mail.tsinghua.edu.cn}

\author[orcid=0000-0001-5284-8001]{Xuefei Chen}
\affiliation{Yunnan Observatories, Chinese Academy of Sciences, Kunming 650216, People's Republic of China}
\affiliation{International Centre of Supernovae (ICESUN), Yunnan Key Laboratory of Supernova Research, Kunming 650216, P.R. China}
\affiliation{University of Chinese Academy of Sciences, Beijing 100049, People's Republic of China}
\email{cxf@ynao.ac.cn}

\author[orcid=0000-0001-9204-7778]{Zhanwen Han}
\affiliation{Yunnan Observatories, Chinese Academy of Sciences, Kunming 650216, People's Republic of China}
\affiliation{International Centre of Supernovae (ICESUN), Yunnan Key Laboratory of Supernova Research, Kunming 650216, P.R. China}
\affiliation{University of Chinese Academy of Sciences, Beijing 100049, People's Republic of China}
\email{zhanwenhan@ynao.ac.cn}


\begin{abstract}

The stability of binary mass transfer is a critical problem for binary evolution. We systematically calculate the adiabatic mass-loss model for naked helium stars with masses ranging from 10\,$M_{\odot}$ to 80\,$M_{\odot}$ to study the critical mass ratio ($q_\textrm{crit}$) of Wolf-Rayet binaries. We set up two prescriptions about Wolf-Rayet stellar wind and consider the isotropic re-emission effect during adiabatic mass loss. Results of the critical mass ratio for conserved dynamically unstable mass transfer show that most of the no-wind helium stars on the main sequence (HeMS) have $0.7<q_\textrm{crit}<3.0$ and on the Hertzsprung gap (HeHG) have $1.5<q_\textrm{crit}<27$. With the Wolf-Rayet star wind effect, the $q_\textrm{crit}$ gets lower on a certain evolutionary stage. With the isotropic re-emission effect, the $q_\textrm{crit}$ gets larger for early-evolutionary stage helium stars and lower for late-evolutionary stage helium stars. Based on fully non-conserved mass transfer, the criteria for HeMS stars are $1.0<q_\textrm{crit}<2.8$ and HeHG stars are $1.5<q_\textrm{crit}<5.0$. Compared with the widely used criterion $q_\textrm{crit}=3$ (HeMS) and $q_\textrm{crit}=4$ (HeHG), our result becomes more unstable for the HeMS stars and more stable for the HeHG stars. Our work could be applied to the binary mass transfer stage of massive helium binaries, such as Wolf-Rayet star binaries and high mass X-ray binaries with Wolf-Rayet star companions. It can be applied to the binary population synthesis studies for the formation of special objects, such as double black hole mergers.

\end{abstract}


\keywords{\uat{Binary evolution}{154} --- \uat{Wolf-Rayet stars}{1806} --- \uat{High mass x-ray binary stars}{733} --- \uat{Helium-rich stars}{715} --- \uat{Stellar evolution}{1599} --- \uat{Common envelope evolution}{2154}}


\section{Introduction}
\label{sec-intro}

The binary fraction of massive stars is considerably impressive among all kinds of stars \citep{2012Sci...337..444S,2017ApJS..230...15M,2024PrPNP.13404083C}. As a result, massive stars have a large chance to fill their Roche lobe during their evolution and overflow to the secondary star. From the perspective of stellar structure and evolution, it is much less certain than for low- and intermediate-mass stars. Influenced by mechanisms such as metallicity \citep{2015MNRAS.452.1068C}, overshooting \citep{2019ApJ...870...77L,2021MNRAS.503.4208S,2023ApJS..268...51L}, rotation \citep{2012A&A...537A.146E,2003A&A...404..975M}, and stellar winds \citep{2022IAUS..366...21S}, stellar structure can vary significantly at all evolutionary stages. It is very possible to lead to a different result during the binary mass transfer stage.

One of the most violent evolutionary stages for massive stars is after they strip all the hydrogen envelope and expose their helium core by strong stellar wind \citep{2003A&A...404..975M}, binary mass transfer \citep{1994A&A...287..803M,2001MNRAS.324...18B,2008A&A...485..245G}, and binary merge \citep{2024ApJ...969..160L}. Because the helium core is produced by convective burning in the core of a massive main-sequence star (MS), the spectrum of stripped envelope stars shows strong helium, nitrogen, and carbon lines \citep{2007ARA&A..45..177C}. From observation, these kinds of stars are defined as Wolf-Rayet stars (WR, \citealt{1867CRAS...65..292W}). WRs are located on the left side of the MS branch on the Hertzsprung–Russell diagram (HRD), which shows that WR stars have a higher surface temperature than MS stars. Based on the surface elemental abundance, WR stars could be separated into different groups \citep{2007ARA&A..45..177C}: WNL (surface hydrogen fraction larger than zero), WNE (no hydrogen detected on the stellar surface), WC (carbon lines dominate on the spectrum), WO (oxygen lines dominate on the spectrum). Among which, the WN is the collection of WNL and WNE, and it has a high probability to be the progenitor of type IIb/Ib supernova \citep{2009ARA&A..47...63S}. One noteworthy feature is that the binary fraction of WR stars is significantly lower than that of MS stars, at only about 40\% \citep{2001NewAR..45..135V}, indicating that a considerable number of companion stars have not been found in WR binaries.

A simple method for simulating WN-type stars is to evolve a star with no hydrogen on the surface, i.e., a naked helium star \citep{2015MNRAS.451.2123T,2016RAA....16..141Y,2018MNRAS.481.1908K,2019ApJ...878...49W}. These stars have a helium-burning core at their centers and a helium envelope. The theory gives a good luminosity-mass relation for WR stars, and the masses for most of the WR observations are in a mass range from $10\,M_{\odot}$ to $25\,M_{\odot}$ \citep{1981ApJ...246..153M,2007ARA&A..45..177C}. It is possible to have the more massive WR stars, but it could be very hard to observe due to the distribution of the initial mass function \citep{1955ApJ...121..161S,2001MNRAS.322..231K}. In this work, we consider the helium star mass larger than $10\,M_{\odot}$ as the massive helium star.

Observations of WR stars show a very strong stellar wind that could reach $\dot{M}_\textrm{wind}\sim10^{-5}\,M_{\odot}/\textrm{yr}$ \citep{1988A&AS...72..259D,1995A&A...299..151H,2000A&A...360..227N,2017A&A...607L...8V,2017MNRAS.470.3970Y,2020MNRAS.499..873S}. Such strong winds could easily create emission lines in the spectrum, and the whole evolution of a helium star could be affected by them. Meanwhile, such strong winds often appear as the helium star mass reaches $8\,M_{\odot}$.  As we introduced
before, the strong wind mass loss rate can affect the simulation of helium stars. In this work, we simply consider a function of WR stellar wind given by \citet{2000A&A...360..227N}, which is determined by luminosity, surface helium fraction, and metallicity of the star. Such strong wind could carry enough orbital angular momentum of the binary system. Depending on the binary's mass ratio, stellar winds can shrink or expand the binary separation.  

Once the WR star in the binary system fills its Roche lobe, the mass will transfer to the accretor star \citep{2020RAA....20..161H}. During this mass transfer stage, the timescale decides the evolution mechanism and the final outcome \citep{1997A&A...327..620S}. Stable mass transfer supposes that the mass transfer timescale is longer than the thermal timescale (most frequently used Kelvin–Helmholtz timescale $\tau_\textrm{{KH}}$) of the donor star. In that case, the thermal energy inside the donor star shall have enough time to maintain the equilibrium, and the donor star can keep the radius near the Roche lobe. On the other hand, a very short timescale for mass transfer shall lead to an adiabatic process, force the violent expansion of the donor star, and finally trigger the common envelope (CE) evolution stage \citep{1976IAUS...73...75P,1978A&A....62..317S}. 

For WR binaries, a stable mass transfer system is more likely to correspond to a high-mass X-ray binary (HMXB) if it has a neutron star (NS) or a black hole (BH) component \citep{2023pbse.book.....T}. If mass transfer is unstable, it is probably very hard to observe because of the short time scale. However, it could result in a very short orbital period, which can contribute to the formation of gravitational wave (GW) sources when the compact objects merge. When a helium star initiates binary mass transfer during core burning, it is classified as Case BA Roche-lobe overflow. After the core burning phase, it is defined as Case BB.

It is crucial to determine the criteria of dynamically unstable mass transfer for different donors. The adiabatic mass loss of polytropic stellar models systematically provides results for such criteria \citep{1987ApJ...318..794H,1997A&A...327..620S}. It simplifies the stellar structure and is widely used in binary population synthesis (e.g., \citealt{2002MNRAS.329..897H,2014A&A...563A..83C}). However, the polytropic model does not seem adequate for late-evolutionary stage stars, given its simple assumption of a fully ionized atmosphere and the ideal gas equation of state. A more accurate method is to use the adiabatic assumption for different initial realistic stellar models and calculate the adiabatic response. The detailed developed processes of adiabatic mass loss are introduced in our previous work of this series of articles: Paper I \citep{PaperI}, II \citep{PaperII}, III \citep{PaperIII}, IV \citep{PaperIV}, and V \citep{PaperV}.

\begin{figure*}[ht!]
	\plotone{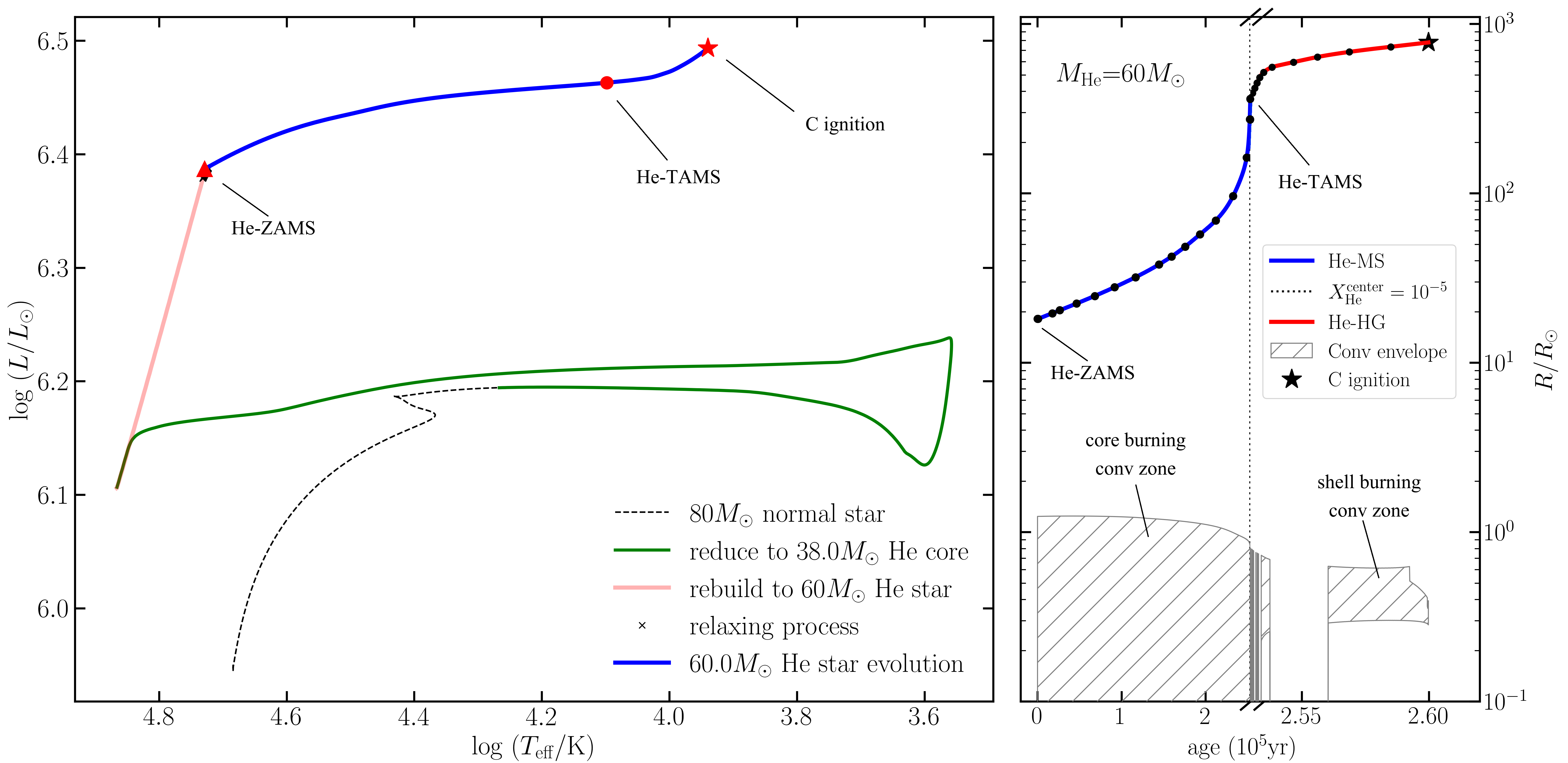}
	\caption{Evolutionary track of a naked $60\,M_{\odot}$ helium star with no wind. In this Paper, 'log' represents the base-10 logarithm. The left panel shows the full evolutionary track from the beginning of HRD, and the right panel shows star age versus stellar radius after He-ZAMS. For the left panel, the black dashed line represents an $80\,M_{\odot}$normal hydrogen envelope star evolving after the MS. The green line is the process of stripping the hydrogen envelope. The pink line shows the mass rebuilding process from a $38\,M_{\odot}$ pure helium core to a $60\,M_{\odot}$ helium star. The blue line represents the evolution stage after He-ZAMS. More detailed discussions of these prescriptions are presented in Section \ref{no wind models}. In the right panel, the solid lines show the surface radii at different evolutionary stages of the helium star. The shadow area represents the convective zone. The black dots indicate the models selected for the adiabatic mass-loss calculations.
    \label{60M HR and R-t}}
\end{figure*}

This research is the continuation of our previous article (Paper IV, \citealt{PaperIV}). We expand the top mass range of helium stars and investigate the mass-transfer stability of massive helium stars. In Section \ref{sec:sequences}, we introduce the massive helium star sequences with different masses and wind schemes. We particularly analyze the impact of the strong wind on decorticating the stellar surface. In Section \ref{sec:adiab}, we use the adiabatic mass-loss model to calculate the mass-loss response of different massive helium stars, based on our sequences and the critical mass-ratio results. Finally, in Section \ref{subsec:discussion}, we compare our results with those of the polytropic model and with observations of WR binaries.

\section{build massive helium star sequences} \label{sec:sequences}

In this Section, we introduce the evolution of massive helium stars. In this Paper, we use Nugis \& Lamers wind \citep{2000A&A...360..227N} prescription to simulate the mass loss rate for massive helium stars. The function of the stellar wind is shown as follows:
\begin{equation}
	\frac{\dot{M}_\textrm{wind} }{M_\odot/\textrm{yr}} = \eta \times  1.0\times 10^{-11} \left ( \frac{L}{L_\odot }  \right ) ^{1.29} \left (X_{\textrm{He}}^{\textrm{surf}}\right )^{1.7} Z^{0.5},
\end{equation}
where $\dot{M}_\textrm{wind}$ is the wind mass loss rate in $M_{\odot}/\textrm{yr}$ unit; $\eta$ is the free parameter; $L$ is the luminosity of the helium star in $L_{\odot}$ unit; $X_{\textrm{He}}^{\textrm{surf}}$ is the surface helium mass fraction and $Z$ is the stellar metallicity.

To calculate the adiabatic mass loss of massive helium stars and the stability criteria for binary mass transfer, we first evolved two sequences of helium star models with different masses: one without stellar winds, and another with $\eta=0.8$ Nugis \& Lamers wind as a suitable prediction of none-rotating massive helium stars \citep{2001A&A...373..555M}. Each star evolves to the core carbon ignition. We selected representative helium-star models at each evolutionary stage to investigate the relationship between binary mass-transfer stability criteria and the nature of different massive helium stars. Meanwhile, Section \ref{wind compression} details the \textit{Wind-driven Decortication Effect}, which could ultimately influence the determination of stability criteria. 

\subsection{Helium Star Sequences without Wind} 
\begin{figure*}[ht!]
	\plotone{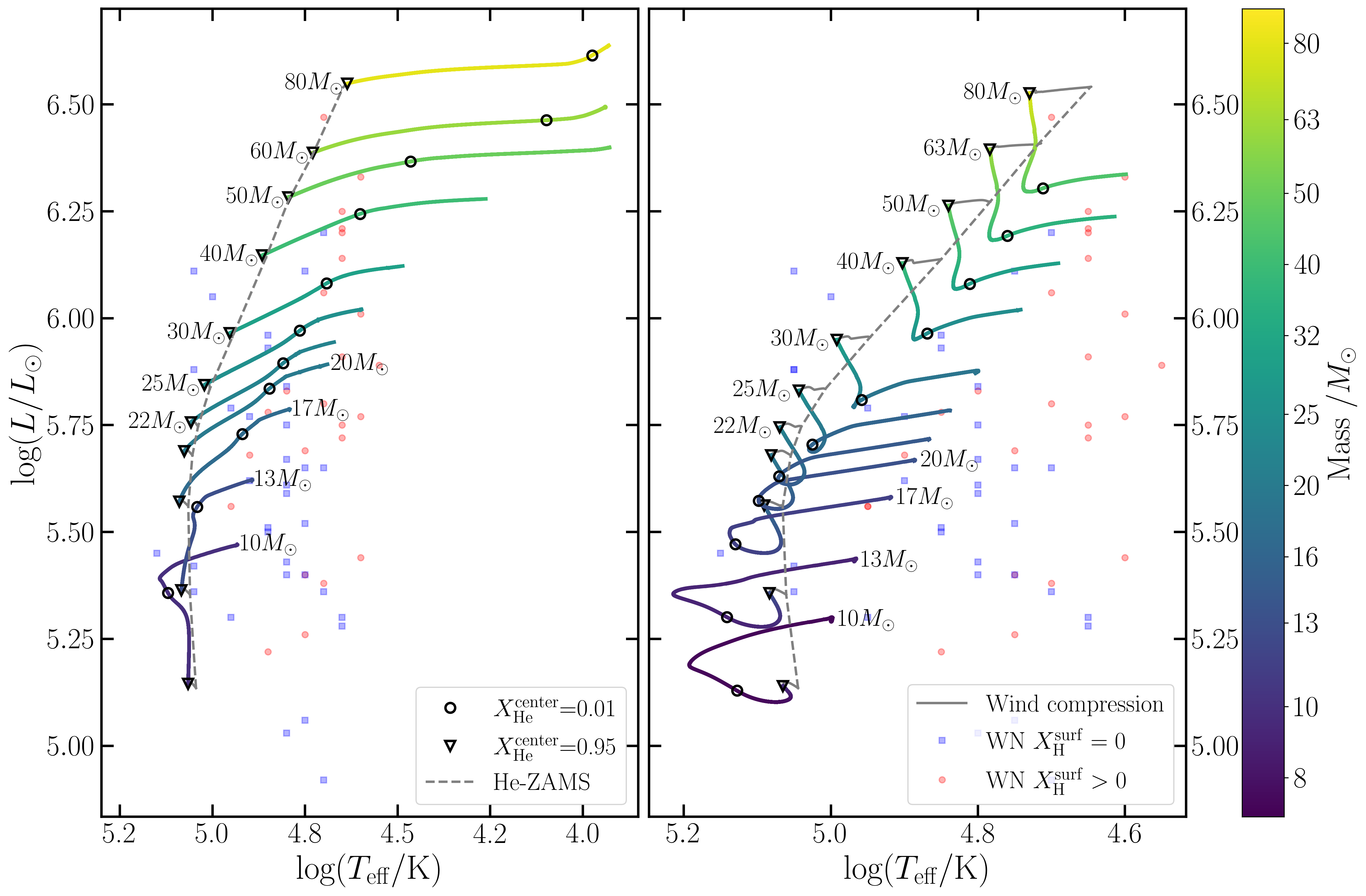}
	\caption{Evolutionary track of naked helium stars on the HR diagram. The left panel shows the models without wind, and the right panel contains 0.8 times Nugis \& Lamers wind. Both sequences are evolved from $10 - 80\,M_{\odot}$ He-ZAMS stars. The color represents the changes in mass during stellar evolution. Due to the no-wind prescription on the left panel, the color is constant. The grey dashed lines in both panels are the initial He-ZAMS models with a central helium fraction of 0.98. Triangle markers represent the center helium fraction that reaches 0.95 on HeMS, which indicates that these stars cross the \textit{Wind-driven Decortication Effect} near the stellar surface and reache the real HeMS (see details in Section\,\ref{wind compression}). Circle markers represent the center helium fraction that reaches 0.01, which we define as He-TAMS. The blue squares and red circles are the observations of WNE and WNL single stars from the galactic observation. \citep{2019A&A...625A..57H}. We find that naked non-rotated helium stars seem to suit the observations relatively well, but can not explain the WRs with a lower effective temperature. 
    \label{HR-wind}}
\end{figure*}

\label{no wind models}

As in our previous research, we use \textit{STARS} code \citet{1971MNRAS.151..351E,1972MNRAS.156..361E,1973MNRAS.163..279E} to build and evolve massive helium stars. It is a one-dimensional non-Lagrangian code. We have already introduced this code in the second Section of Paper I \citep{PaperI}. Using this code, we evolved a grid of non-rotating, wind-free massive helium-star models across different masses.

Following the tradition, we adopted a metallicity of $Z=0.02$ for our simulations of Population I stars. The overshooting parameter $\delta$=0.12 (\citealt{1997MNRAS.285..696S,1998MNRAS.298..525P}) and the mixing length parameter $\alpha$=2.0 \citep{1998MNRAS.298..525P} are also calibrated values. The present study does not explore the evolution of massive helium stars in metal-poor environments.

As shown in Figure\,\ref{60M HR and R-t}, we artificially stripped the hydrogen envelope to create helium stars. Similar to Paper IV \citep{PaperIV}, all helium stars were constructed through the following steps: Firstly, We evolve a massive MS star (from $50\,M_{\odot}$ to $90\,M_{\odot}$) to the phase after terminal age of main-sequence (TAMS) and before core helium ignition, where the star possesses a non-degenerate helium core and a hydrogen envelope. Secondly, we artificially stop the nuclear abundance variations for all elements heavier than hydrogen to preserve the helium core structure. Meanwhile, we artificially strip the hydrogen envelope material by using a strong wind several times larger than the Nugis \& Lamers prescription until the hydrogen of the star is completely depleted. Thirdly, stop the wind mass loss and reopen the abundance variation. After the process of relaxation, the star successfully becomes a zero-age main-sequence helium star (He-ZAMS). Finally, under the no-wind condition, we evolved the helium stars until the ignition of elements heavier than helium (typically core carbon burning) occurred. 

\begin{figure*}[ht!]
	\plotone{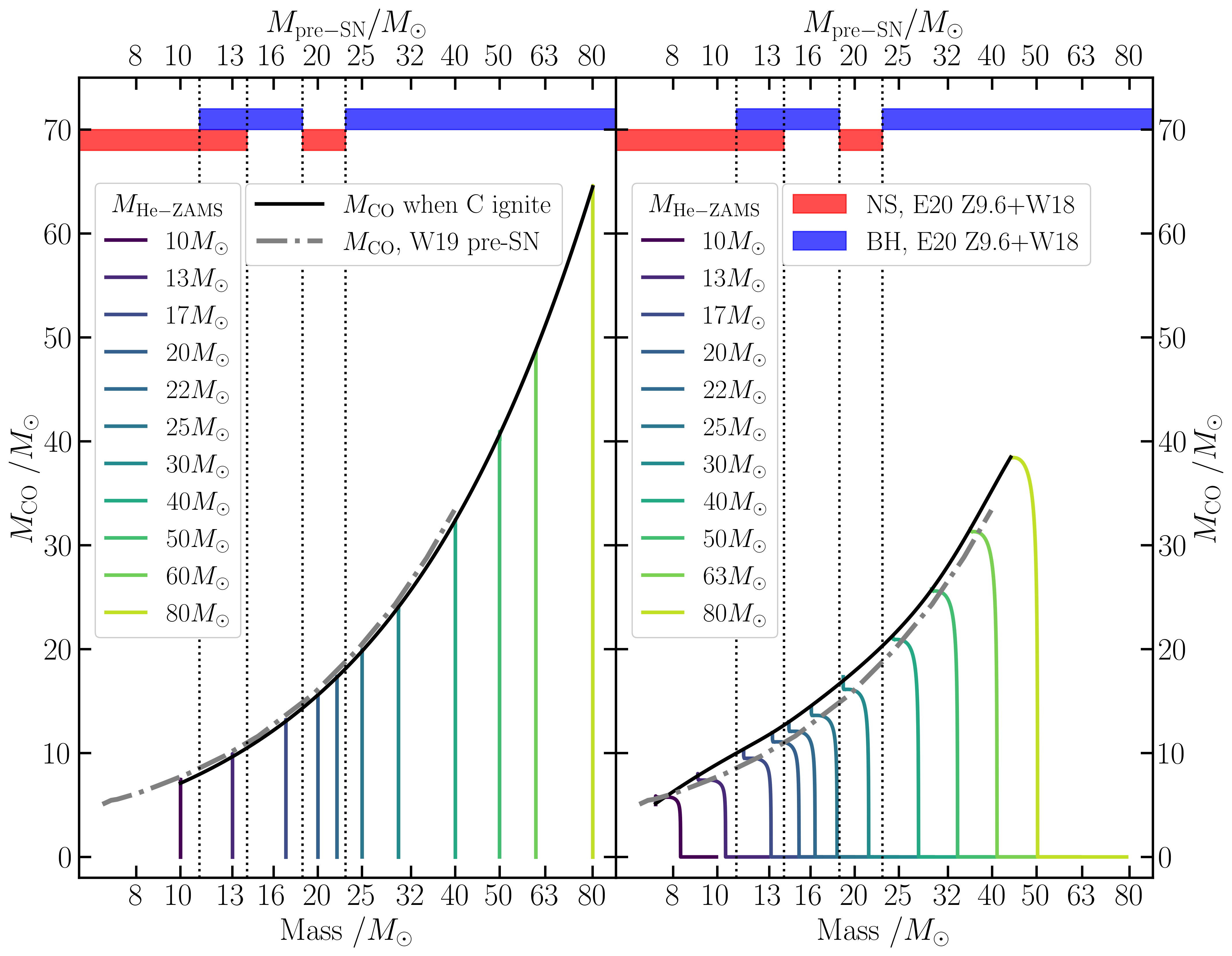}
	\caption{$M_\textrm{CO}$ of helium star sequences. The x-axis is the helium star mass, and the y-axis is $M_\textrm{CO}$ for this specific model. Similar to Figure\,\ref{HR-wind}, the left panel shows the models without wind, and the right panel shows $\eta=0.8$ Nugis \& Lamers wind models. Due to the no-wind prescription on the left, the CO mass evolutionary tracks are vertical lines, whereas the total stellar masses decrease in the right panel.
	The grey dashed-dotted lines are the pre-supernova models from \citealt{2019ApJ...878...49W}, which use a combined WR wind from \citealt{2017MNRAS.470.3970Y} (the standard wind scheme in their work). The blue and red bands show the remnant types after the supernova explosion and the fallback from \citealt{2020ApJ...890...51E} with the Z9.6+W18 neutrino engine. The mass ranges for NS and BH are from the pre-supernova models.
		\label{Mco-wind}}
\end{figure*}

We terminate the onset of carbon ignition when the heavy
element nuclear burning luminosity reaches $100\,L_\odot$. After the onset of core carbon burning, the star has approximately 1000 years of remaining lifetime, which is relatively short compared with the unstable binary mass-transfer timescale \citep{1978A&A....62..317S}. So we ignore the possibility of binary mass transfer after carbon ignition in this research.

Typically, a $100\,M_{\odot}$ main-sequence star can produce a helium star of $40\sim50\,M_{\odot}$. Given that the upper mass limit for helium stars may exceed $50\,M_{\odot}$, once the star reaches the target mass in the third stage of building a helium star, we artificially increase helium-rich material on the stellar surface by applying a negative wind prescription, ultimately enabling the helium star masses up to $80\,M_{\odot}$. Due to numerical instabilities encountered in some stellar models, the final constructed no-wind helium star mass sequence consists of [10, 13, 17, 20, 22, 25, 30, 40, 50, 60, 80]$\,M_{\odot}$. These masses roughly have equal intervals on a log scale.

At the beginning of He-ZAMS, the center helium mass fraction of all stars is equal to 0.98 for every helium star. Subsequently, stable helium burning ignites in the stellar core, establishing a convective burning region at the center. Throughout most of the HeMS phase, the triple-alpha process (3$\alpha$ reaction) dominates core helium burning. When the central helium abundance $X_\textrm{He}^\textrm{cntr}$ drops below around 0.1, the 4$\alpha$ reaction (combination of the carbon-$\alpha$ reaction plus the 3$\alpha$ reaction) becomes the primary nuclear process, converting a fraction of carbon in the center into oxygen and ultimately forming a non-degenerate carbon-oxygen (CO) core inside the helium star. When $X_\textrm{He}^\textrm{cntr}$ drops below about $10^{-2}$, center helium burning stops, and we define the massive helium star as having reached the terminal age of the helium main sequence (He-TAMS).

Compared with low-mass HeMS stars, the nuclear reactions of massive HeMS stars and structures are similar, but the trends in stellar radius evolution differ significantly. For low-mass helium stars, we find a shrink phase during the 4$\alpha$-reaction-dominated stage. However, for the helium stars heavier than $17\,M_{\odot}$, the radius of the surface  keeps increasing
during the HeMS. We believe the constantly expanding stage near He-TAMS for more massive stars is influenced by the shell burning convective zone (like the right panel of Figure\,\ref{60M HR and R-t} shows). For the helium star larger than $17\,M_{\odot}$, the maximum radius on HeMS occurs on He-TAMS.

After HeMS, helium stars stop core helium burning and enter the shell-burning phase. In this stage, the helium stars have a CO core and a radiative helium-rich envelope. The structure at this stage is very similar to that of massive normal stars evolving into the Hertzsprung gap. Here, we call such massive helium Hertzsprung-gap stars (HeHG). The evolution of HeHG occurs over a very short timescale, but it can effectively increase the radius of helium stars. Similar to the trend of low- and intermediate-mass helium stars we previously studied in Paper IV \citep{PaperIV}, massive helium stars with $M>10\,M_{\odot}$ will ignite carbon before they evolve to a total convective helium envelope. Therefore, massive helium stars do not have a helium-giant branch (HeGB).

\begin{figure*}[ht!]
	\plotone{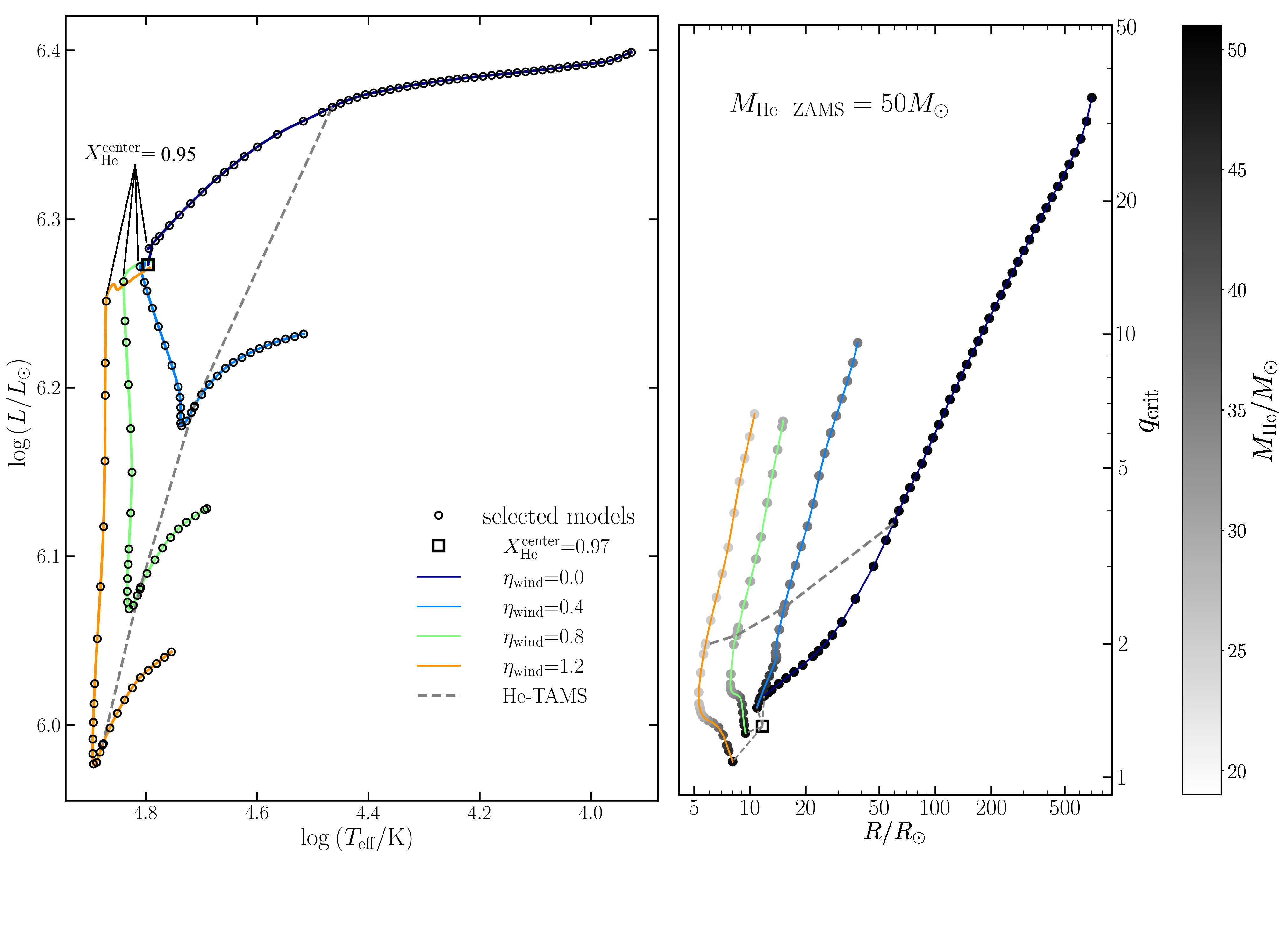}
	\caption{The evolutionary track on the HR diagram (left) and the conserved critical mass ratio (right) for the selected models with different wind prescriptions. The grey dashed lines represent the models that reach He-TAMS. Both four evolution curves start from one $50\,M_{\odot}$ He-ZAMS model (the square marker). With different wind schemes, massive helium stars evolve to different masses, as shown by the colors of the dots in the right panel.
		\label{50M-wind}}
\end{figure*}

Because the lifetime after carbon ignition could be shorter than $1000\,\textrm{yr}$, helium stars without a dramatic expansion are unlikely to undergo a common envelope evolution. We suspect that the possibility of this part of helium stars participating in binary star evolution is relatively low. Therefore, we ignore the parameter space after carbon ignition. Massive helium stars are composed of a non-degenerate CO core and a helium-rich envelope. Compared with the CO core mass $M_\textrm{CO}$ of the pre-supernova models by \citet{2019ApJ...878...49W}, our results are very similar.
We anticipate that they are most likely to evolve to type IIb or type Ib supernovae (SN IIb/Ib, \citealt{2009ARA&A..47...63S}). In Figure\,\ref{Mco-wind}, we compare our CO core masses with the simulation results of SN IIb explosions \citep{2020ApJ...890...51E}. In their work, supernova explosion simulations indicated that SN IIb could form both NSs and BHs, depending on the compactness of the helium star models before core collapse. The mass range for the different outcomes (NS and BH) is shown in Figure\,\ref{Mco-wind}. Due to the 
uncertainties of core collapse mechanisms and neutrino engines, it is not realistic to predict the detailed remnant mass, but the trend for different types of remnant stars is still valuable. Here, the mass range is provided for reference only.

\subsection{Helium Star Sequences with WR Wind } 
\label{Nugis wind}

Although we did not consider stellar wind effects during stellar evolution in our previous papers, stellar winds are among the key physical processes that impact stellar structure. When it comes to massive stars, especially for WR-type stars, ignoring the wind effect makes the results less reliable. We want to compare the two sequences, with and without stellar wind, to observe differences in their structures and critical mass ratios at the same mass and radius.

As we have talked at the beginning of this Section, we consider the Nugis \& Lamers wind as a suitable prescription for massive helium stars. Since we have not introduced the effect of stellar rotation, the scale factor of Nugis and Lamers
wind $\eta = 0.8$ is set in our WR wind helium star sequence. 

We introduce the helium stellar wind after the star evolves past the He-ZAMS. Due to the same reason of numerical instabilities, the final constructed WR-wind helium star mass sequence consists of [10, 13, 17, 20, 22, 25, 30, 40, 50, 63, 80]$\,M_{\odot}$. Stellar wind mass loss rate is around $10^{-5}\sim10^{-3}\,M_{\odot}/\textrm{yr}$, which is comparable to the observation of WR-type stellar wind. Helium stars with stellar winds also evolve to carbon ignition, and the evolutionary tracks are shown in the right panel of Figure\,\ref{HR-wind}. On HeMS, stellar wind mass loss can effectively strip the envelope and compress the central convection zone mass. On He-TAMS, the total mass loss fraction of a helium star is close to $40\%$ compared to He-ZAMS. Therefore, the stellar radius undergoes a shrinkage stage during HeMS. After He-TAMS, due to a very short lifetime on HeHG, the stellar wind could not strip considerable mass, and the expansion trend on HeHG is not affected.

Due to the mass-loss rate of massive helium stars, the helium envelope is continually stripped from the star, and the CO core mass fraction increases at later evolutionary stages. The $M_\textrm{CO}$ during the helium star evolution with stellar wind is in the right panel of Figure\,\ref{Mco-wind}. The solid lines show that the $M_\textrm{CO}$ fraction is larger if using the WR stellar wind prescription. Meanwhile, the incomplete ionization zone near the surface is shrinking, and its radius is shrinking further due to wind effects. We will discuss such \textit{Wind-driven Decortication Effect} in the next Section \ref{wind compression}. 

In our work, we select some helium-star models to calculate adiabatic mass loss. On HeMS, we use $X_\textrm{He}^\textrm{cntr}$ as the principle, and on HeHG, we use certain equal intervals in a log scale of radius variation. The changes of stellar structure could affect the binary mass transfer stability, which we will discuss in Section \ref{sec:adiab}.

\subsection{Wind-driven Decortication Effect} 
\label{wind compression}

Normally, the stellar wind does not directly influence the stellar structure. Like MS stars, the stellar wind mass loss rate is too slow to influence the thermal equilibrium of stars. However, for massive helium stars, the mass loss rate could be larger than $10^{-5} \,M_{\odot}/\textrm{yr}$, which is very likely to influence the local thermal equilibrium near the surface.

\begin{figure}[ht!]
	\plotone{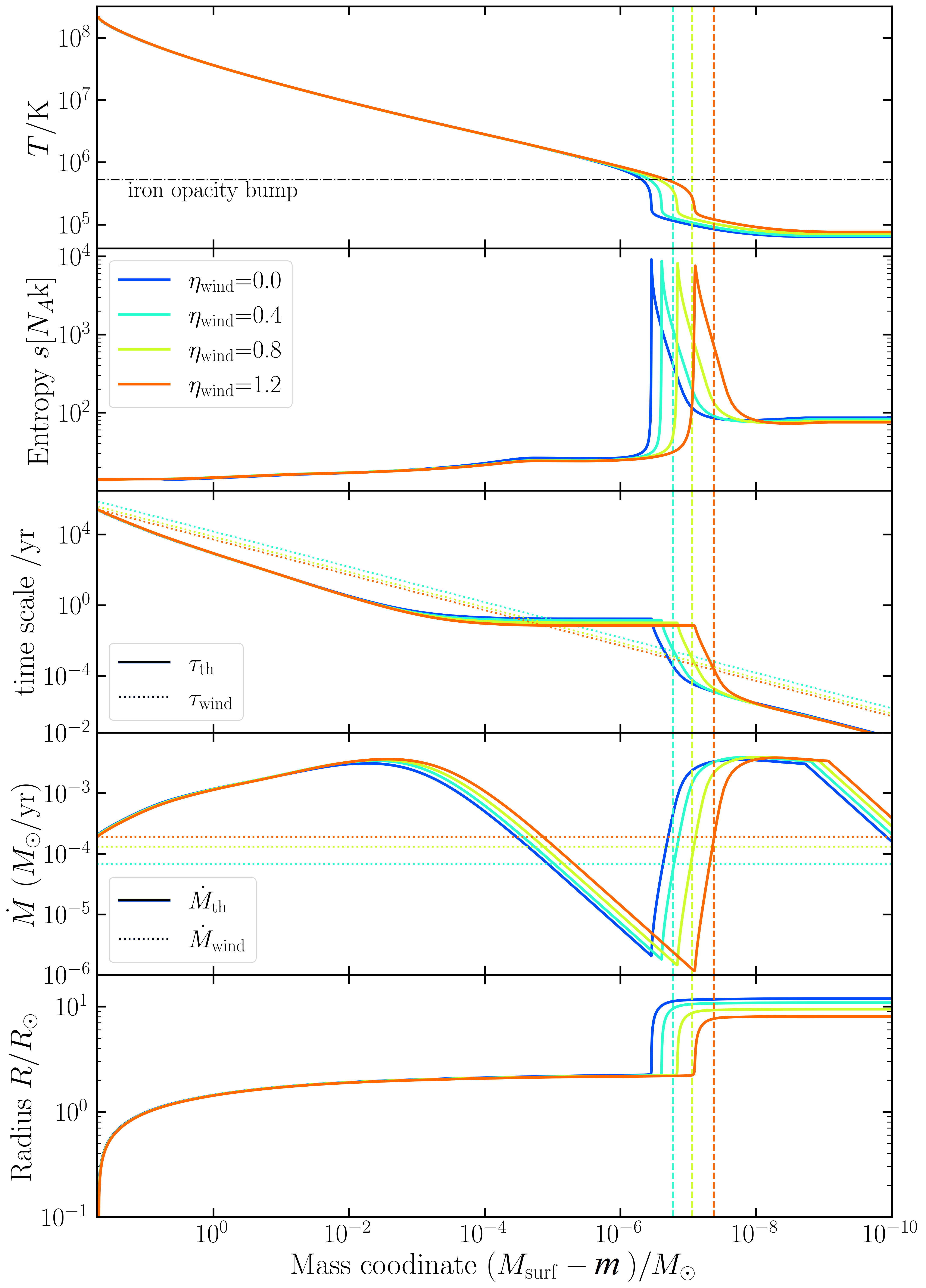}
	\caption{The structures of different helium stars with different stellar winds. The x-axis is the mass coordinate from the surface to the center (where $m$ is the mass coordinate). The four stars evolve from one $50\,M_{\odot}$ He-ZAMS model and share the same center helium fraction of $X^\textrm{center}_\textrm{He}=0.95$. From the top panel to the bottom, the curves represent the profiles of temperature, entropy, $\tau _\textrm{th}$, $\dot{M} _\textrm{th,crit}$ and the radius. In the top panel, the iron opacity bump occurs at $T = 10^{5.3}\textrm{K}$. The vertical lines are the boundaries once the stellar wind reaches the $\dot{M} _\textrm{th,crit}$. 
    \label{50-0.95wind-compression}}
\end{figure}

The effect of massive helium wind appears near the He-ZAMS with the increase of stellar mass. As shown in the right panel of Figure\,\ref{HR-wind}, at the beginning of HeMS, the massive helium stars firstly evolve to the high temperature side in a very short timescale (the grey lines) and the radii of these stars are compressed. This feature could only be induced by stellar wind. Then the evolutionary tracks turn to another tendency. The inflection points near the helium stars with a $X_\textrm{He}^\textrm{cntr}$ of 0.95.

To show the details of this process, we simulate an extra grid of $50\,M_{\odot}$ helium stars with different wind factors $\eta = 0, 0.4, 0.8, 1.2$. The left panel of Figure\,\ref{50M-wind} shows the evolutionary track of these stars on the HR diagram. As the wind mass-loss rate increases, the radius compression becomes more significant at the beginning of HeMS. We believe the stellar structure is directly disturbed by the wind, and that this mechanism should operate at other evolutionary stages as well. Unfortunately, with mass changes and varying stellar ages, stellar structure can change, making it more difficult to distinguish the effect from stellar winds. Therefore, we pick four models at the inflection points near He-ZAMS ($X^\textrm{center}_\textrm{He}=0.95$ models, which are pointed out in Figure\,\ref{50M-wind}) to avoid those effects. These four models have the same initial mass and evolutionary stage, pass through the inflection points, and still have a total mass loss that is not significant. Here we show the structure of these stars in Figure\,\ref{50-0.95wind-compression}.

In this Figure, we plot stellar structures as a function of mass. Stellar winds are not strong enough to influence the inner structure, so the central physics of these stars is quite similar. However, the structures near the photosphere are decorticated. In this Figure, we use a log scale for the mass coordinate from the surface to highlight the surface area. The panel at the top shows the temperature profiles. With increased wind mass loss, the area of rapid temperature changes is closer to the surface. Compared with the iron opacity bump temperature \citep{1992ApJ...397..717I,2006A&A...450..219P}, such a structure represents the envelope inflation structure (EIS, e.g. \citealt{2012A&A...538A..40G,2013ApJS..208....4P,2023A&A...674A.216L}) of the helium star. This area can also be observed in the entropy profile (second panel) as the surface superadiabatic region, corresponding to the right side of the peak. As we can see, the increase in the massive helium stellar wind leads to a shrinkage of the surface EIS area, decorticates the outer boundary of EIS and effectively compresses the total radius. Similar results also appear in \citealt{2016ApJ...821..109R}.

The strong wind mass loss rate forces the surface thermal structure out of thermal equilibrium. Here, we use the thermal local timescale to show this mechanism. This timescale represents the time for complete loss of internal energy in the stellar outer envelope using stellar luminosity as the power, which can be calculated as follows \citep{2013sse..book.....K},
\begin{equation}
	\tau _\textrm{th} \left ( m \right ) =\frac{1}{L}\int_{m}^{M_\textrm{surf}}  C_{p} \left ( {m}'  \right ) T \left ( {m}'  \right ) \mathrm{d}{m}'.  
\end{equation}
In this Equation, $m$ represents the mass coordinate, $M_\textrm{surf}$ represents the total mass, $L$ represents stellar luminosity, $ C_{p}$ represents the heat capacity at constant pressure at the specific mass coordinate, and $T$ is the temperature. The local thermal timescales for different mass-loss rates are shown as the solid lines in the third panel of Figure\,\ref{50-0.95wind-compression}. To reach the limit of local thermal equilibrium, the average mass loss rate of a certain mass coordinate $\dot{M} _\textrm{th}$ should reach a critical limit \citep{2023A&A...669A..45T}:
\begin{equation}
	\dot{M} _\textrm{th,crit} \left ( m \right ) =\frac{M_\textrm{surf}-m }{\tau _\textrm{th} \left ( m \right ) }. 
\end{equation}
Such critical thermal mass loss rates are shown in the fourth panel. In this panel, the horizontal dot lines represent the specific mass-loss rate of helium stellar winds. Once the wind mass loss rate is higher than $\dot{M} _\textrm{th,crit}$, the thermal equilibrium from this layer to the surface is broken. Therefore, the thermal relaxation process on the left side of the vertical dot lines is delayed. Due to wind-induced mass stripping, the mass near the EIS region is reduced, though the inner structure remains. As a result, the radius profile (the panel at the bottom) shows that the strong stellar wind compresses the radius structures near the surface. We define such a mechanism as the \textit{Wind-driven Decortication Effect}.

Compared with all helium stars, for He-ZAMS models, with increasing initial helium stellar mass, the depth at which the wind mass-loss rate exceeds $\dot{M}_\textrm{th}$ gradually becomes apparent (the detailed information is shown in Appendix B). As a result, the \textit{Wind-driven Decortication Effect} gets more significant, and the grey solid lines in the right panel of Figure\,\ref{HR-wind} are getting longer after He-ZAMS. 
\textit{Wind-driven Decortication Effect} should also appear in all the evolutionary stages once the wind mass loss rate is higher than $\dot{M}_\textrm{th,crit}$. In Appendix B, we have calculated the $\dot{M}_\textrm{th,crit}$ for no-wind models and compared with the typical wind mass-loss rate, $\eta = 1.0$. Stellar radius may decrease with the development of wind prescription at all stages of helium star evolution. We have noticed that some researchers artificially enlarged the convective boundaries to avoid the occurrence of the region near the Eddington limit (e.g., MLT++ in MESA code \citealt{2013ApJS..208....4P}), which will lead to a reduction in the radius of the star \citep{2023A&A...674A.216L}. In this work, we have not considered such a method.

The importance of the stellar \textit{Wind-driven Decortication Effect} is that the stellar radius strongly influences the critical mass ratio before adiabatic mass loss. In addition, other uncertainties may also affect the radius of helium stars, such as the convective overshooting parameter \citep{2016RAA....16..141Y} and the rotation \citep{2003A&A...404..975M}. This article will not discuss them further. In the next Section, we will present the details of the adiabatic mass-loss calculation for massive helium stars.

\section{Adiabatic Mass Loss} \label{sec:adiab}

As described in Section \ref{sec:sequences}, we already provided the prescription for choosing representative models for adiabatic mass loss. In this part, we discuss how to calculate the mass-transfer stability criteria for massive helium stars and the detailed distribution of the critical mass ratio in the Mass-Radius ($M-R$) parameter space. Such a mechanism has already been introduced in our previous works.

\subsection{Calculation of Stability Criterion} 
\label{subsec:qcrit}

Generally speaking, we use the adiabatic assumption to simulate the dynamical timescale mass-loss response in massive helium stars. In our model, the WR star is the donor (with stellar mass of $M_\textrm{He}$) and the type of the accretor (with stellar mass of $M_\textrm{acc}$) is not limited. In this process, the entropy profile follows the initial structure of the initial donor star model. We assume the helium star is the donor star and the mass ratio of the binary system is $q=\frac{M_\textrm{He}}{M_\textrm{acc}}$. In the adiabatic process, we focus on the evolution of the helium star's radius. Meanwhile, using a certain mechanism for orbital angular momentum loss, we can calculate the Roche lobe radius of the donor star.

\begin{figure*}[ht!]
	\plotone{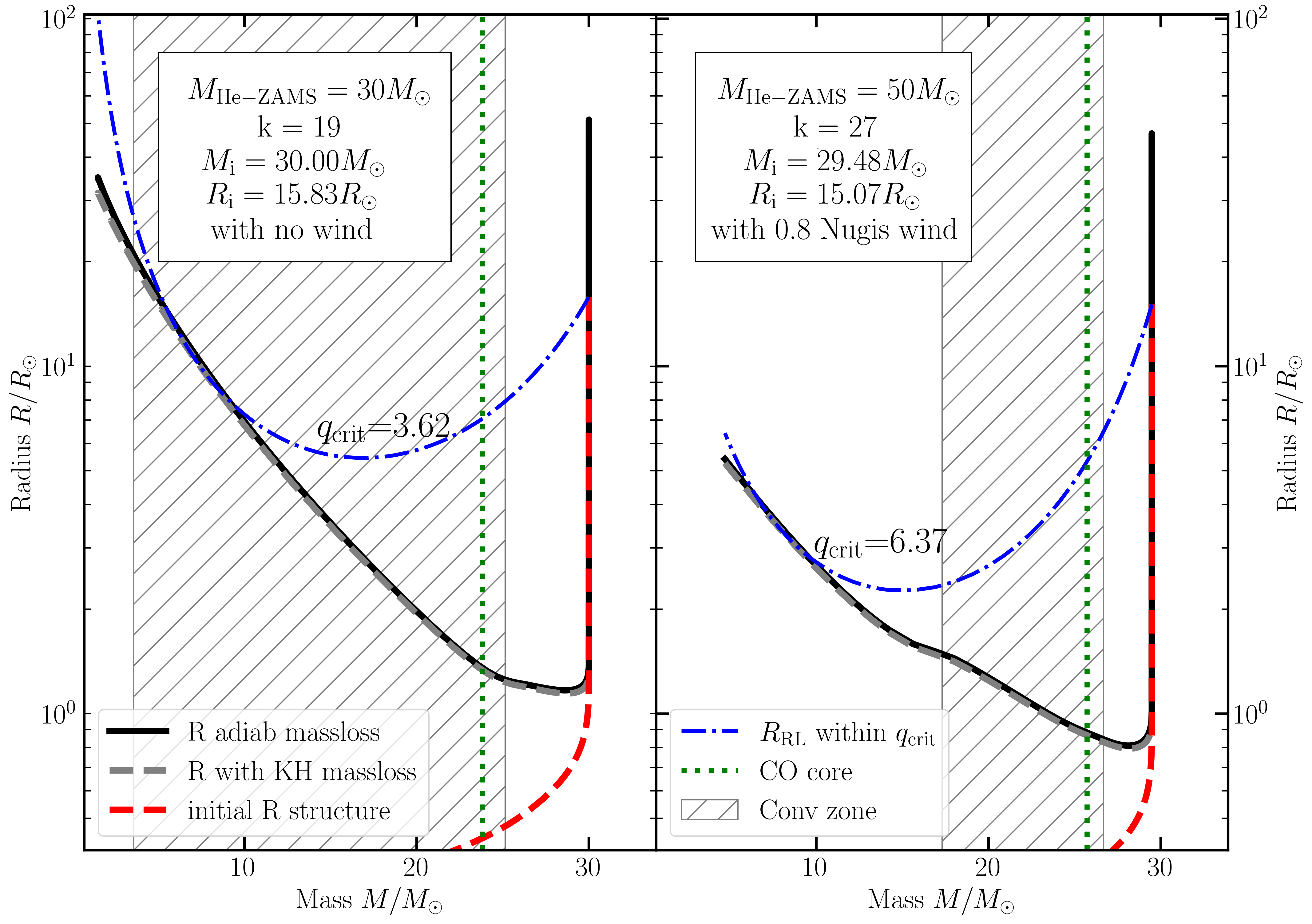}
	\caption{The radius evolution during the adiabatic mass loss of two massive helium stars. They share similar mass and radius but in different wind schemes. The left panel shows a no-wind model with $30\,M_{\odot}$, which just evolves to the early HeHG stage. The right panel shows a model with the $\eta=0.8$ wind scheme and an initial mass of $50\,M_{\odot}$, which has evolved to the late HeHG stage. The black lines are the surface radius, and the blue dashed lines are the Roche lobe radius when the initial mass ratio equals the conserved $q_\textrm{crit}$. Due to different initial structures and entropy profiles, the results of the $q_\textrm{crit}$ are quite different.
		\label{compare_MR}}
\end{figure*}

During the adiabatic mass loss process of the massive helium star, we construct a layer inside the star to make the adiabatic mass loss rate equal to KH timescale mass loss rate when the Roche lobe radius reaches this layer (detailed description is in Paper I \citealt{PaperI}). We define the radius of this layer as $R_\textrm{RH}$. For a critical binary mass transfer situation, the Roche lobe radius is supposed to be just above the $R_\textrm{RH}$ during all mass loss stages. Using this method, we calculate the stability criteria for massive helium stars. The detailed simulation follows some assumptions below:
\begin{itemize}
	\item [1)]
	The initial Roche lobe radius equals the helium stellar radius before adiabatic mass loss, which is the trigger of binary mass transfer. We assume that adiabatic mass loss begins at the onset of binary mass transfer.
	\item [2)]
	During the adiabatic mass-loss process, the angular momentum loss in the binary orbit is dominated by the mass loss of the binary system. Using a mass-loss mechanism, the Roche-lobe radius change can be determined from the initial binary mass ratio before the adiabatic simulation.
	\item [3)]
	The secondary star is always smaller than its Roche lobe. Therefore, the criteria are not limited to the secondary type. We assume the binary system remains tidally locked during mass loss and that no spin angular momentum is transferred to orbital angular momentum.
	\item [4)]
	Before the Roche lobe radius reaches $R_\textrm{RH}$, the star should go through a delayed unstable mass transfer stage. We ignore the possible thermal-equilibrium response of this stage.
	\item [5)]
	We use the initial mass ratio as the criterion. For a critical situation, the Roche lobe radius should always be larger than $R_\textrm{RH}$ and the certain layer where $R_\textrm{RL}=R_\textrm{RH}$ is the maximum depth for delayed unstable mass transfer.
\end{itemize}
In this work, we consider the mass loss mechanism of the binary system as the isotropic re-emission \citep{1973A&A....25..387V,1991PhR...203....1B} of the secondary star. In this pattern, the mass is ejected near the accretor, and the fraction of mass loss $\beta$ is a free parameter. So we have
\begin{equation}
	\dot{M}_\textrm{acc} = (1-\beta) \dot{M}_\textrm{donor},
\end{equation}
where the $\dot{M}_\textrm{donor}$ is the mass loss rate from the donor star and the $\dot{M}_\textrm{acc}$ is the mass accretion rate from the accretor. The detailed angular momentum loss of isotropic re-emission is described in Paper V \citep{PaperV}.

\begin{figure*}[ht!]
	\plotone{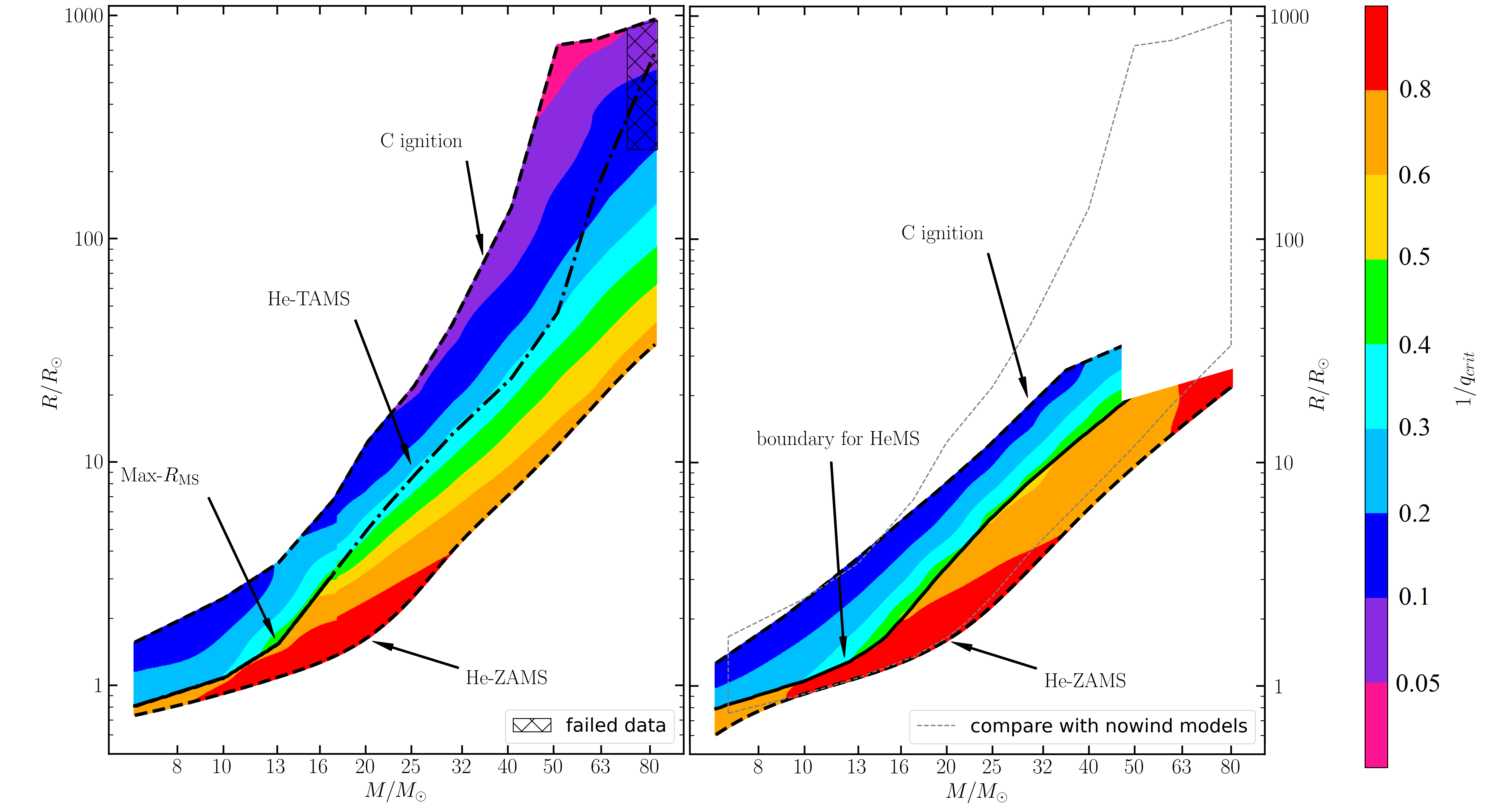}
	\caption{The results of $q_\textrm{crit}$ for the conserved mass transfer prescription in the $M-R$ parameter space. The left panel is for the no-wind scheme, and the right one is for the $\eta=0.8$ scheme. The grey-dashed area on the right panel is the same space as the one to the left. The maximum field of the chosen helium star models sets the boundary for HeMS in the right panel. Due to the \textit{Wind-driven Decortication Effect}, the radii of He-ZAMS shrink with the development of the stellar wind, which becomes more significant with the increase of stellar mass.
		\label{MR-space}}
\end{figure*}

Figure\,\ref{compare_MR} gives two samples of the radius evolution during adiabatic mass loss. Both models share similar mass and radius but differ in structure and evolutionary stage, driven by different wind prescriptions. Red dashed lines are the radius profile in the mass coordinate of initial helium star models. The Roche lobe radii of the two panels represent the critical situation for the conserved mass transfer $\beta=0$. Under the wind effect, the adiabatic mass-loss theory yields very different values of $q_\textrm{crit}$ for the two similar helium stars.

The possible mass ratio range for a helium binary system is $q>0$. If the initial mass ratio is larger than $q_\textrm{crit}$, the Roche lobe radius gets smaller and the star shall go through dynamical unstable mass transfer. Otherwise, the binary system is supposed to keep thermal equilibrium during binary mass transfer. As shown near the surface, the stellar radii of both stars expand at the onset of adiabatic mass loss. This is caused by a very thin convection zone near the initial model surface, and the entropy peak also occurs there. A similar behavior also occurs in hydrogen-rich, radiative-envelope massive stars.

We have shown that the different wind prescriptions could lead to similar stellar mass and radius but very different $q_\textrm{crit}$. In the right panel of Figure\,\ref{50M-wind}, we give the evolution of $q_\textrm{crit}$ for the same He-ZAMS mass models. From He-ZAMS to the end of evolution, the $q_\textrm{crit}$ of massive helium stars keeps increasing. With the increase of wind factor, the maximum values of $q_\textrm{crit}$ decrease, and the trend of increasing criterion with radius change becomes steeper. 

Compared with other models, it is questionable whether radius is a reasonable parameter for quantifying a star's evolutionary stage. Before He-TAMS, the center fraction of helium was a good parameter, but on the HeHG, it is very hard to find a single norm to represent the evolution stage. CO core mass fraction and normalized age may be effective parameters but they are too model-dependent. Considering the importance of stellar radius as a trigger of binary mass transfer, we still use the radius as the symbol of stellar evolution in this article.

\subsection{Mass-transfer Stability Criteria In M-R Parameter Spaces} 
\label{subsec:qcrit}

In this Section, we put every massive helium star model in the $M-R$ parameter space. The interpolate method in $M-R$ space is introduced in Appendix C, and it is similar to what we did in Paper IV \citep{PaperIV}. In Figure\,\ref{MR-space} we give the conserved results ($\beta=0$) of no-wind models on the left panel and $\eta=0.8$ models on the right panel. The color in this Figure represents the result of the $q_\textrm{crit}$. If the $q$ of one system is larger than $q_{\textrm{crit}}$ at the beginning of mass transfer, the system should go through the dynamical unstable mass transfer phase. For the most common situation, $q_{\textrm{crit}}$ is unlikely to reach a value over 10 because of the initial binary mass ratio distribution. To highlight the lower value of the binary mass ratio, we use $1/q_{\textrm{crit}}$ as the label of the counter map. All the results of $q_{\textrm{crit}}$ are based on the interpolation in Appendix C.

For the models of the failed data area in the left panel, we cannot obtain good results because of rapid surface shrinkage during adiabatic mass loss. A similar problem appears in lower-mass helium stars and asymptotic-branch giant stars in our previous papers. The grey-dashed area on the right of the Figure is the same space as the one to the left. Limited by the maximum initial mass, the $M-R$ parameter space of massive HeHG stars with $\eta=0.8$ wind scheme is missing. With the wind effect, the radii are lower than in the no-wind parameter space.   

As the $q_{\textrm{crit}}$ on $M-R$ space shows, the maximum $q_{\textrm{crit}}$ decreases at a certain evolutionary stage with the influence of massive helium star wind. Comparing two systems with similar mass and radius, the evolutionary stage for a star is later than another star with no wind, so that the $q_{\textrm{crit}}$ could be larger, like Figure\,\ref{compare_MR}. In general, the $q_{\textrm{crit}}$ is influenced by stellar mass, evolutionary stage (represented by stellar radius), and the stellar wind scheme. In our grid, none of the $q_{\textrm{crit}}$ reach the limitation of Darwin instability (DI,  \citealt{1879RSPS...29..168D,1995ApJ...444L..41R,2001ApJ...562.1012E,2019arXiv190701877S}), which requires the critical DI mass ratio $q_\textrm{DI}$ over 100 (for example, the minimum $q_\textrm{DI}\sim200$ for $50\,M_{\odot}$ helium stars in our calculation).

Though there is still some weakness in using the stellar radius as a symbol for evolutionary stage, we predict that the $M-R$ parameter space for $q_{\textrm{crit}}$ is the most suitable and convenient mass transfer criteria table for binary population synthesis (BPS). In the next Section, we will briefly discuss the improvement in BPS using our results and compare them with possible observations for massive helium-star binary systems.

\begin{figure*}[ht!]
	\plotone{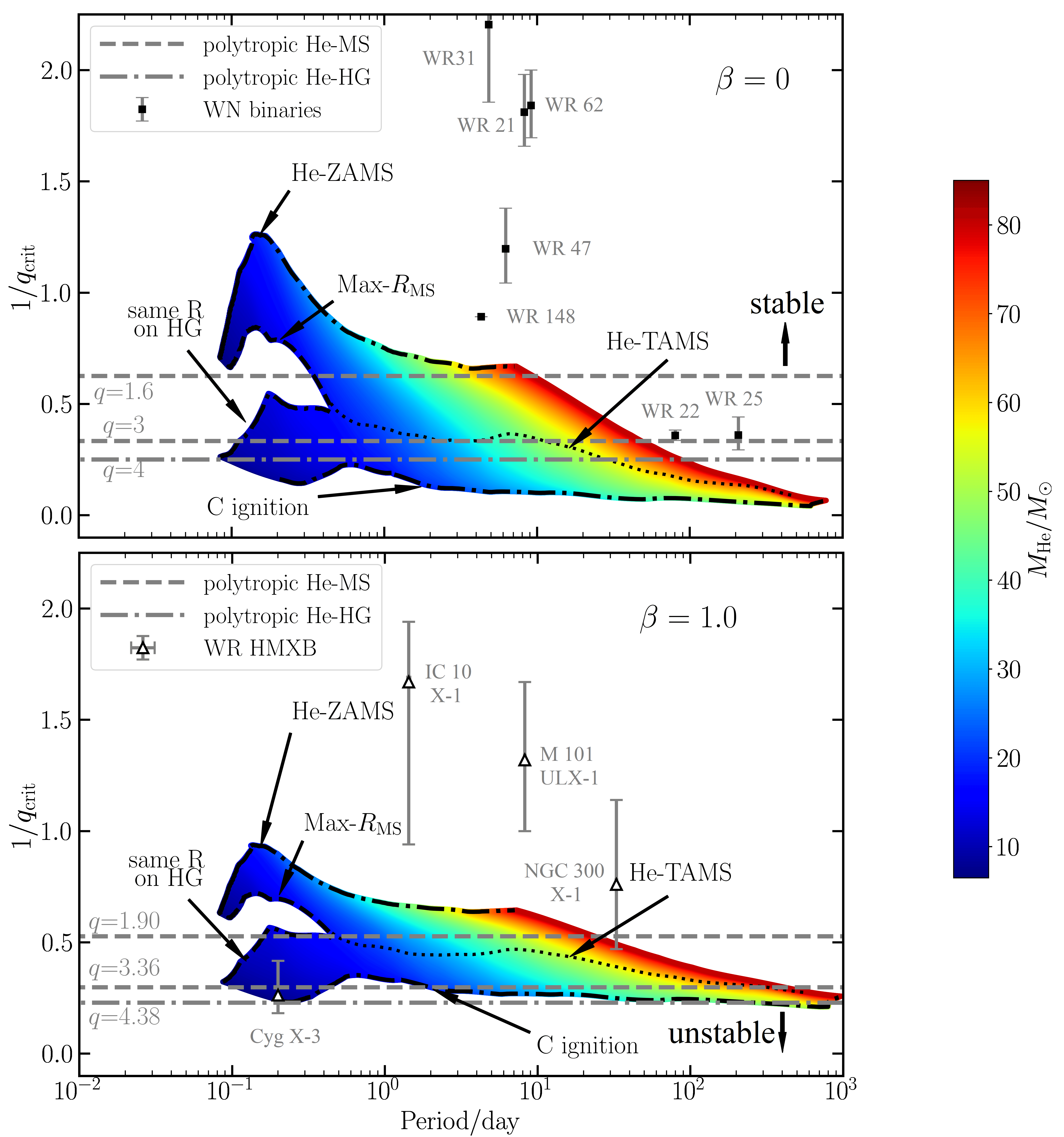}
	\caption{The distribution of critical mass ratio on Mass-Radius in period-mass ratio parameter space. The panels are based on different treatments of angular momentum loss. The top panel is conserved mass transfer, and the bottom panel is the completely non-conserved mass transfer of isotropic re-emission angular momentum loss ($\beta=1$). The colorbar represents the stellar mass. The horizontal lines are the elder criterion used in BPS (see details in Appendix D). We plot the known WN-type binary stars and the WR HMXB systems. The angular momentum loss during mass transfer is conserved for an MS star but extremely non-conserved for a compact accretor. Here we use conserved $q_\textrm{crit}$ on the top panel to compare with WN+O systems, and $\beta=1$ scheme on the bottom panel to compare with the WR HMXBs. Detailed information is described in Sections \ref{subsec:WN+O} and \ref{subsec:WN+HMXB}. 
		\label{Pq-space}}
\end{figure*}

\section{Discussion} 
\label{subsec:discussion}

For the binary stars, it is considerably harder to observe the mass and radius of both stars directly. On the contrary, with the detection of the spectrum and light curve, it is more accurate and uncomplicated to measure the binary orbital period ($P_\textrm{orb}$) and mass ratio. Here we shift our $q_{\textrm{crit}}$ parameter space into $P_\textrm{orb}-q$ space to compare with some representative helium binary objects, as Figure\,\ref{Pq-space} shows. The method of calculating the $P_\textrm{orb}$ was introduced in previous work Paper IV \citep{PaperIV}. We will discuss these objects in the following subsections.

In general, the adiabatic mass model shows that $q_{\textrm{crit}}$ increases from He-ZAMS to carbon ignition in massive helium stars. Most of the conserved $q_{\textrm{crit}}$ is limited within the range of 0.7 to 3 for HeMS stars. For HeHG stars, the conserved $q_{\textrm{crit}}$ increases from 1.5 to 27 with the evolution of the star. Such a result is very close to the low- and intermediate-mass helium stars in Paper IV \citep{PaperIV}. Based on the summarized data on total helium stars, we predict that mass transfer in HeHG is extremely stable, consistent with the Case BB mass transfer predictions \citep{2015MNRAS.451.2123T,2017ApJ...846..170T,2018MNRAS.481.4009V,2019MNRAS.490.3740N}, which is necessary for double NS systems.

If we consider the $\eta=0.8$ wind within the helium star evolution, the results of $q_{\textrm{crit}}$ get lower for the helium stars with a similar evolutionary stage, which suggests a more unstable prospect for WR wind models. If we use a completely non-conserved isotropic re-emission mass transfer prescription  ($\beta=1$) for no wind models, the $q_{\textrm{crit}}$ on HeMS is limited within the range of 1.0 to 2.8, and the maximum $q_{\textrm{crit}}$ on HeHG decreases to 5. 

In Figure\,\ref{Pq-space}, the top panel gives the no-wind $q_{\textrm{crit}}$ with conserved angular momentum loss. Compared with the conserved polytropic criterion, which are $q_{\textrm{crit}}=1.6$ \citep{2014A&A...563A..83C,2007A&A...467.1181D} or $q_{\textrm{crit}}=3$ \citep{2023A&A...669A..82L,2002MNRAS.329..897H} for HeMS and $q_{\textrm{crit}}=4$ for HeHG \citep{2002MNRAS.329..897H,1997MNRAS.291..732T}, the adiabatic models predict that the parameter space of unstable mass transfer for HeMS is wider, while for HeHG it is narrower. This result is similar to the results for low- and intermediate-mass helium stars. 

The bottom panel is the no-wind $q_{\textrm{crit}}$ $\beta=1$ prescription. In this panel, the $q_{\textrm{crit}}$ gets lower for the early-evolutionary stage helium stars, and the $q_{\textrm{crit}}$ gets larger for the late-evolutionary stage stars. For a more reasonable comparison, we estimate the $\beta=1$ polytropic criterion based on the conserved criterion above, which is detailed described in Appendix D. Due to the $\beta$ having a tight connection with the accretion star, conserved $q_{\textrm{crit}}$ fits the non-compact object well while $\beta=1$ is more suitable for BH. Here, we use the WN+O systems and HMXBs with WR companions to constrain our results.

\subsection{Compare with WN+O Type Binaries} 
\label{subsec:WN+O}

From Gaia and other observations, several WR binary systems were found in the Milky Way \citep{2019A&A...625A..57H}. These binary systems have an O-type companion, and most of these WR primaries are WN-type stars. Based on the luminosity-mass relations of WR stars \citep{2012ARA&A..50..107L,2011A&A...535A..56G}, the orbital periods and mass ratios are given by several researches \citep{2012MNRAS.424.1601F,1996A&A...306..771R,2008RMxAC..33...91G,2017MNRAS.467.3105M,2013A&A...552A..22C}. We plot these observations on the top panel of Figure\,\ref{Pq-space}.

Based on the observation of these WR binary systems, there is no evidence of binary mass transfer, i.e., Roche lobe overflow. From the perspective of mass transfer, it might not be appropriate to compare these systems with our critical mass ratio. However, the known systems are all stable. From this point of view, only a stable system could survive if they have to go through the mass transfer stage.

\subsection{Compare with HMXBs with a WR Companion} 
\label{subsec:WN+HMXB}

Some of HXMBs have a WR companion (WR HMXB). The most well-known object is Cyg X-3 \citep{1973A&A....25..387V, 1996A&A...314..521V}, which is a very luminous and periodic X-ray source located at Cygnus \citep{2012MNRAS.426.1031Z}. The infrared spectrum observation shows that Cyg X-3 has HeI and HeII emission lines, which are the special features of the WN type star. The radial velocity period from the spectrum shows a synchronized pace with the X-ray light curve during the low state \citep{2009A&A...501..679V}. It is believed that the WR and the X-ray source are in a single binary system, with material transferred to the accretor. It is believed that Cyg X-3 has a BH accretor \citep{2010ApJ...718..488S} and the mass ratio $q=3.8_{-1.4}^{+1.7}$ \citep{2009A&A...501..679V}.

For now, there is still only one WR HMXB system, Cyg X-3, found in the Milky Way, and three more WR HMXB systems, IC 10 X-1 \citep{2007ApJ...669L..21P}, M 101 X-1 \citep{2013Natur.503..500L}, NGC 300 X-1 \citep{2010MNRAS.403L..41C}, were detected in other galaxies. These observations are plotted on the bottom panel of Figure\,\ref{Pq-space} to compare with the $\beta=1$ parameter space. Based on their radial velocities, the mass ratios have been studied over the last two decades. In the bottom panel of Figure\,\ref{Pq-space}, we plot these systems and compare them with our criteria. IC 10 X-1, M 101 X-1, and NGC 300 X-1 are in a stable mass-transfer stage, consistent with our predictions. Other WR HXMB systems, such as CG X-1 \citep{2004ApJ...605..360W,2015MNRAS.452.1112E}, lack sufficient observations to confirm their mass ratios. However, we noticed that Cyg X-3 is a unique object that is extremely close to the criteria limit.

For all the WR HMXB systems, researchers assume the WR stars are He-ZAMS and use the mass-luminosity relation \citep{1989A&A...210...93L} to estimate the mass of WR. Based on the prediction of the WR mass of Cyg X-3: $10\,M_{\odot}<M_\textrm{WR}<40\,M_{\odot}$, the mass ratio $q=3.8_{-1.4}^{+1.7}$ of Cyg X-3 is neither stable for $\beta=1$ polytropic criteria of HeMS $q=3.36$ nor our HeMS criterion (lower than the dotted line in the bottom panel of Figure\,\ref{Pq-space}). Considering the mass loss from the donor star, the initial binary mass ratio at the beginning of the mass transfer may be even larger. If we add the error of mass ratio, the polytropic criterion HeMS $q=3.36$ has the possibility to satisfy the stable mass transfer stage for Cyg X-3. For the most common situation, due to the long-term X-ray emission, the WR HMXB systems should maintain stable mass transfer. But we find it is difficult to explain by both the HeMS binary mass transfer criterion.

We propose the following possibilities:
Firstly, if the WR in Cyg X-3 is a HeHG star, both of the criteria above could satisfy the stable mass transfer.In this situation, the mass-luminosity relation is no longer applicable, which introduces bias for the mass ratio of Cyg X-3. If we ignore such bias, based on our parameter space of the stable mass transfer, the top limit of WR mass is $8\,M_{\odot}$ for $\beta=1$ and $11\,M_{\odot}$ for $\beta=0$.
Secondly, the observation of Cyg X-3 predicted that the WR radius is larger than its Roche lobe radius. It is also possible that Cyg X-3 is undergoing a delayed unstable mass-transfer stage but hasn't reached CE evolution. A similar object is SS433, an A-type supergiant companion HMXB system \citep{2017MNRAS.471.4256V}.
Thirdly, the system is not at the beginning of Case BA/BB mass transfer but at the end phase of Case B mass transfer (the donor star is a HG or RGB star). The system has finished the unstable mass transfer phase, lost its hydrogen envelope, and the primary star became a WN-type star. Before the mass transfer stops, the system changes to a temporary stable mass transfer phase (or so-called CE decoupling phase) and forms the source that we have observed \citep{2025ApJ...979..112N}.

The mass of WR has a great deal of uncertainty. From the perspective of verifying the stable mass transfer criterion, Cyg X-3 is likely a valuable system for testing the physics near the dynamical mass transfer. It would be important to detect systems like Cyg X-3 to assess the accuracy of the different binary mass-transfer criteria. We hope to have other evidence in the future to prove its mass transfer status.

\section{Summary} \label{summary}

In this research, we have calculated the adiabatic mass loss of massive helium stars with initial masses in the range $10\,M_{\odot}<M_\textrm{WR}<80\,M_{\odot}$. With different prescriptions for stellar wind, we have studied mass sequences with no-wind stars and $\eta=0.8$ of Nugis \& Lamers' WR wind. Based on the different structures of helium stars, we found that strong winds can significantly decrease stellar radii, a phenomenon we term the \textit{Wind-driven Decortication Effect} in this work. We systematically build full massive helium star models from He-ZAMS to the core carbon ignition.

We use the adiabatic mass loss code to calculate the adiabatic response of these models. Meanwhile, with different prescriptions for isotropic re-emission of angular momentum loss, we calculate the Roche lobe radius during mass loss. From the definition of stable criteria $R_\textrm{RL} \geq R_\textrm{RH}$, we determined the critical mass ratio of unstable mass transfer for the massive helium stars. For no-wind and conserved prescription, we found most of our models follow $0.7<q_\textrm{crit}<3.0$ on HeMS and $1.5<q_\textrm{crit}<27$ on HeHG. With the WR wind effect, the $q_\textrm{crit}$ gets lower on a certain evolutionary stage. With the isotropic re-emission effect, the $q_\textrm{crit}$ gets larger for early-evolutionary stage helium stars and lower for late-evolutionary stage helium stars. Take a example of $\beta=1$ no-wind models, $1.0<q_\textrm{crit}<2.8$ on HeMS and $1.5<q_\textrm{crit}<5.0$ on HeHG. Detailed parameter space is shown in the Figure\,\ref{MR-space} and Figure\,\ref{Pq-space}.

WR observations during the binary mass-transfer stage are rare. Compared with WR+O binaries and WR HMXBs, most observations in our work are consistent with a stable mass-transfer scenario. However, Cyg X-3 is very close to our stability criterion. It may be a very important object to be studied in the future.

Combining with low- and intermediate-mass helium stars in Paper IV \citep{PaperIV}, we predict that more binary systems undergo unstable mass transfer in Case BA than the polytropic model suggested, and the Cass BB is likely to be extremely stable. In the next Paper, we will present the fitting formula for the critical mass ratios and investigate the detailed differences by applying different $q_\textrm{crit}$ values for helium binaries to binary population synthesis studies. So far, we have not investigated the total metallicity space of helium stars. We will further expand the $q_\textrm{crit}$ parameter space of helium stars into the metal-poor and extreme metal-poor stars like \citealt{2023ApJ...945....7G} and \citealt{PaperV}.

\begin{acknowledgments}
	We thank the anonymous referee for comments that help to improve the article. This project is supported by the Strategic Priority Research Program of the Chinese Academy of Sciences (grant No. XDB1160201), National Natural Science Foundation of China (NSFC, grant Nos. 12288102, 12525304, 12125303, 12173081, 12333008, 12422305), National Key R\&D Program of China (grant Nos. 2021YFA1600403, 2021YFA1600401), Yunnan Revitalization Talent Support Program - Science \& Technology Champion Project (No. 202305AB350003), Yunnan Fundamental Research Projects (No. 202401BC070007), International Centre of Supernovae, Yunnan Key Laboratory (No. 202302AN360001), the CAS "Light of West China" and the Young Talent Project of Yunnan Revitalization Talent Support Program, the New Cornerstone Science Foundation through the XPLORER PRIZE. LZ thanks all the collaborators for their selfless help.
\end{acknowledgments}

\appendix

\section{Physical parameters and mass transfer stability criteria of helium stars}

Here, we show the data for the mass sequence $25\,M_{\odot}$ as samples. The Table\,\ref{tab:nowind} is for the no wind models and Table\,\ref{tab:nugis-wind} is for $\eta=0.8$ Nugis \& Lamers wind prescription. The completed mass sequence data are released in a machine-readable table. The columns of Table\,\ref{tab:nowind} and Table\,\ref{tab:nugis-wind} are as follows:

\noindent
1. k --- sequence model number, indicate the order from He-ZAMS to C ignition;\\
2. mass --- the stellar mass of the current model;\\
3. age --- evolution age measured from He-ZAMS model (k=01 model);\\
4. $M_\textrm{CO}$ --- carbon/oxygen core mass in solar unit;\\
5. log $R$ --- initial helium stellar radius before mass loss;\\
6. log $t_\textrm{{KH}}$ --- Kelvin–Helmholtz timescale of the initial model;\\
7. log $L$ --- initial helium stellar luminosity before mass loss;\\
8. log $T_{\rm eff}$ --- initial helium stellar effective temperature before mass loss;\\
9. log $g$ --- the acceleration of gravity on the stellar surface;\\
10. $X_{\textrm{He}}^{\textrm{cntr}}$ --- helium fraction at the center;\\
11. $q_{\textrm{crit}}^0$ --- critical mass ratio for $\beta=0$ conserved angular momentum loss;\\
12. $q_{\textrm{crit}}^{0.5}$ --- critical mass ratio for $\beta=0.5$ half non-conserved angular momentum loss;\\
13. $q_{\textrm{crit}}^1$ --- critical mass ratio for $\beta=1$ fully non-conserved angular momentum loss;\\
14. type --- evolutionary stage of helium star;\\
Note: The '...' marker represents the failed data of $q_{\textrm{crit}}$.

\begin{table}[htbp]
	\centering
	\caption{Physical parameters and mass transfer stability of a no-wind $25\,M_{\odot}$ helium star} 
	\label{tab:nowind}
	\begin{tabular}{cccccccccccccc}
		\hline
		k & mass & age & $M_\textrm{CO}$ & log $R$ & log $t_\textrm{{KH}}$ & log $L$ & log $T_{\rm eff}$ & log $g$ & $X_{\textrm{He}}$ & $q_{\textrm{crit}}^0$ & $q_{\textrm{crit}}^{0.5}$ & $q_{\textrm{crit}}^1$ & type \\ 
		{} & $M_{\odot}$ & yr & $M_{\odot}$ & $R_{\odot}$ & yr & $L_{\odot}$ & ${\rm K}$ & ${\rm cm/s^2}$ & {} \\
		\hline
		01  &  25.00  &  0.00e+00  &  14.41  &  0.9233  &  4.0463  &  5.8437  &  1.6139  &  5.0343  &  0.95065  &   0.941  &   1.075  &   1.188  & HeMS\\ 
		03  &  25.00  &  3.77e+04  &  18.16  &  1.0627  &  3.9725  &  5.8569  &  1.6085  &  4.9132  &  0.80049  &   1.081  &   1.208  &   1.298  \\ 
		05  &  25.00  &  9.39e+04  &  19.27  &  1.2819  &  3.8581  &  5.8762  &  1.5999  &  4.7227  &  0.60048  &   1.307  &   1.413  &   1.453  \\ 
		07  &  25.00  &  1.57e+05  &  19.84  &  1.5245  &  3.7320  &  5.8969  &  1.5902  &  4.5120  &  0.40031  &   1.543  &   1.617  &   1.603  \\ 
		09  &  25.00  &  2.11e+05  &  20.15  &  1.7232  &  3.6284  &  5.9143  &  1.5823  &  4.3395  &  0.25027  &   1.766  &   1.802  &   1.729  \\ 
		11  &  25.00  &  2.54e+05  &  20.34  &  1.8708  &  3.5509  &  5.9276  &  1.5764  &  4.2112  &  0.15028  &   1.962  &   1.960  &  ...  \\ 
		13  &  25.00  &  3.06e+05  &  20.50  &  2.0503  &  3.4547  &  5.9458  &  1.5692  &  4.0553  &  0.05024  &   2.272  &   2.197  &  ...  \\ 
		\hline
		15  &  25.00  &  3.35e+05  &  20.51  &  2.2190  &  3.3611  &  5.9662  &  1.5626  &  3.9088  &  0.00117  &   2.814  &   2.589  &   2.208  & HeHG\\ 
		17  &  25.00  &  3.36e+05  &  20.51  &  2.2454  &  3.3463  &  5.9695  &  1.5616  &  3.8858  &  0.00024  &   2.922  &   2.664  &  ...  \\ 
		19  &  25.00  &  3.36e+05  &  20.52  &  2.3239  &  3.3032  &  5.9785  &  1.5585  &  3.8177  &  0.00000  &   3.271  &   2.898  &  ...  \\ 
		21  &  25.00  &  3.37e+05  &  20.53  &  2.4640  &  3.2291  &  5.9917  &  1.5528  &  3.6960  &  0.00000  &   3.925  &   3.309  &   2.575  \\ 
		23  &  25.00  &  3.41e+05  &  20.57  &  2.6032  &  3.1592  &  6.0013  &  1.5469  &  3.5751  &  0.00000  &   5.679  &   4.334  &   3.026  \\ 
		25  &  25.00  &  3.42e+05  &  20.60  &  2.7414  &  3.0924  &  6.0080  &  1.5408  &  3.4550  &  0.00000  &   6.799  &   4.916  &   3.258  \\ 
		27  &  25.00  &  3.43e+05  &  20.61  &  2.8799  &  3.0265  &  6.0137  &  1.5347  &  3.3347  &  0.00000  &   8.017  &   5.503  &   3.483  \\ 
		29  &  25.00  &  3.44e+05  &  20.58  &  3.0164  &  2.9624  &  6.0187  &  1.5285  &  3.2162  &  0.00000  &   9.799  &   6.313  &   3.760  \\ 
		\hline 
	\end{tabular}
	\tablecomments{Table 1 is published in its entirety in the machine-readable format.\\
		A portion is shown here for guidance regarding its form and content.}
\end{table}
\begin{table}[htbp]
	\centering
	\caption{Physical parameters and mass transfer stability of a $\eta=0.8$ Nugis \& Lamers wind $25\,M_{\odot}$ helium star} 
	\label{tab:nugis-wind}
	\begin{tabular}{cccccccccccccc}
		\hline
		k & mass & age & $M_\textrm{CO}$ & log $R$ & log $t_\textrm{{KH}}$ & log $L$ & log $T_{\rm eff}$ & log $g$ & $X_{\textrm{He}}^{\textrm{cntr}}$ & $q_{\textrm{crit}}^0$ & $q_{\textrm{crit}}^{0.5}$ & $q_{\textrm{crit}}^1$ & type \\ 
		{} & $M_{\odot}$ & yr & $M_{\odot}$ & $R_{\odot}$ & yr & $L_{\odot}$ & ${\rm K}$ & ${\rm cm/s^2}$ & {} \\
		\hline
		01  &  24.78  &  0.00e+00  &  14.24  &  0.8111  &  4.0999  &  5.8310  &  1.6181  &  5.1278  &  0.95077  &   0.816  &   0.951  &   1.081  & HeMS\\ 
		03  &  23.42  &  3.85e+04  &  17.22  &  0.8196  &  4.0716  &  5.8069  &  1.6165  &  5.0961  &  0.80047  &   0.888  &   1.022  &   1.143  \\ 
		05  &  21.58  &  9.58e+04  &  17.16  &  0.8380  &  4.0260  &  5.7732  &  1.6140  &  5.0444  &  0.60065  &   0.998  &   1.129  &   1.232  \\ 
		07  &  19.71  &  1.60e+05  &  16.55  &  0.8543  &  3.9758  &  5.7375  &  1.6115  &  4.9909  &  0.40038  &   1.099  &   1.225  &   1.311  \\ 
		09  &  18.30  &  2.16e+05  &  15.78  &  0.8291  &  3.9491  &  5.7107  &  1.6113  &  4.9805  &  0.25046  &   1.145  &   1.268  &   1.343  \\ 
		11  &  17.67  &  2.47e+05  &  15.30  &  0.8009  &  3.9417  &  5.6999  &  1.6120  &  4.9898  &  0.17955  &   1.153  &   1.274  &   1.350  \\ 
		13  &  17.00  &  2.87e+05  &  14.74  &  0.7639  &  3.9335  &  5.6906  &  1.6131  &  5.0051  &  0.10028  &   1.172  &   1.291  &   1.361  \\ 
		15  &  16.58  &  3.16e+05  &  14.39  &  0.7384  &  3.9260  &  5.6875  &  1.6141  &  5.0165  &  0.05003  &   1.210  &   1.327  &   1.391  \\ 
		\hline 
		17  &  16.18  &  3.49e+05  &  14.04  &  0.7299  &  3.8964  &  5.6995  &  1.6150  &  5.0131  &  0.00109  &   1.468  &   1.558  &   1.567  & HeHG\\ 
		19  &  16.17  &  3.50e+05  &  14.04  &  0.7488  &  3.8830  &  5.7043  &  1.6145  &  4.9966  &  0.00001  &   1.561  &   1.637  &   1.624  \\ 
		21  &  16.15  &  3.52e+05  &  14.03  &  0.8864  &  3.7994  &  5.7271  &  1.6096  &  4.8765  &  0.00000  &   2.088  &   2.068  &   1.917  \\ 
		23  &  16.12  &  3.54e+05  &  14.02  &  1.0260  &  3.7197  &  5.7448  &  1.6044  &  4.7546  &  0.00000  &   2.802  &   2.599  &   2.243  \\ 
		25  &  16.10  &  3.56e+05  &  14.01  &  1.1643  &  3.6484  &  5.7546  &  1.5989  &  4.6338  &  0.00000  &   3.694  &   3.201  &   2.563  \\ 
		27  &  16.09  &  3.57e+05  &  14.03  &  1.3019  &  3.5804  &  5.7621  &  1.5932  &  4.5139  &  0.00000  &   4.383  &   3.631  &   2.764  \\ 
		29  &  16.08  &  3.57e+05  &  14.05  &  1.4403  &  3.5122  &  5.7698  &  1.5875  &  4.3934  &  0.00000  &   5.109  &   4.060  &   2.961  \\ 
		31  &  16.07  &  3.58e+05  &  14.09  &  1.5772  &  3.4447  &  5.7774  &  1.5818  &  4.2744  &  0.00000  &   5.959  &   4.543  &   3.176  \\ 
		33  &  16.07  &  3.58e+05  &  14.37  &  1.6884  &  3.3895  &  5.7842  &  1.5772  &  4.1777  &  0.00000  &   7.728  &   5.479  &   3.564  \\ 
		\hline 
	\end{tabular}
	\tablecomments{Table 2 is published in its entirety in the machine-readable format as well.\\
		A portion is shown here for guidance regarding its form and content.}
\end{table}

\begin{figure*}[ht!]
	\plotone{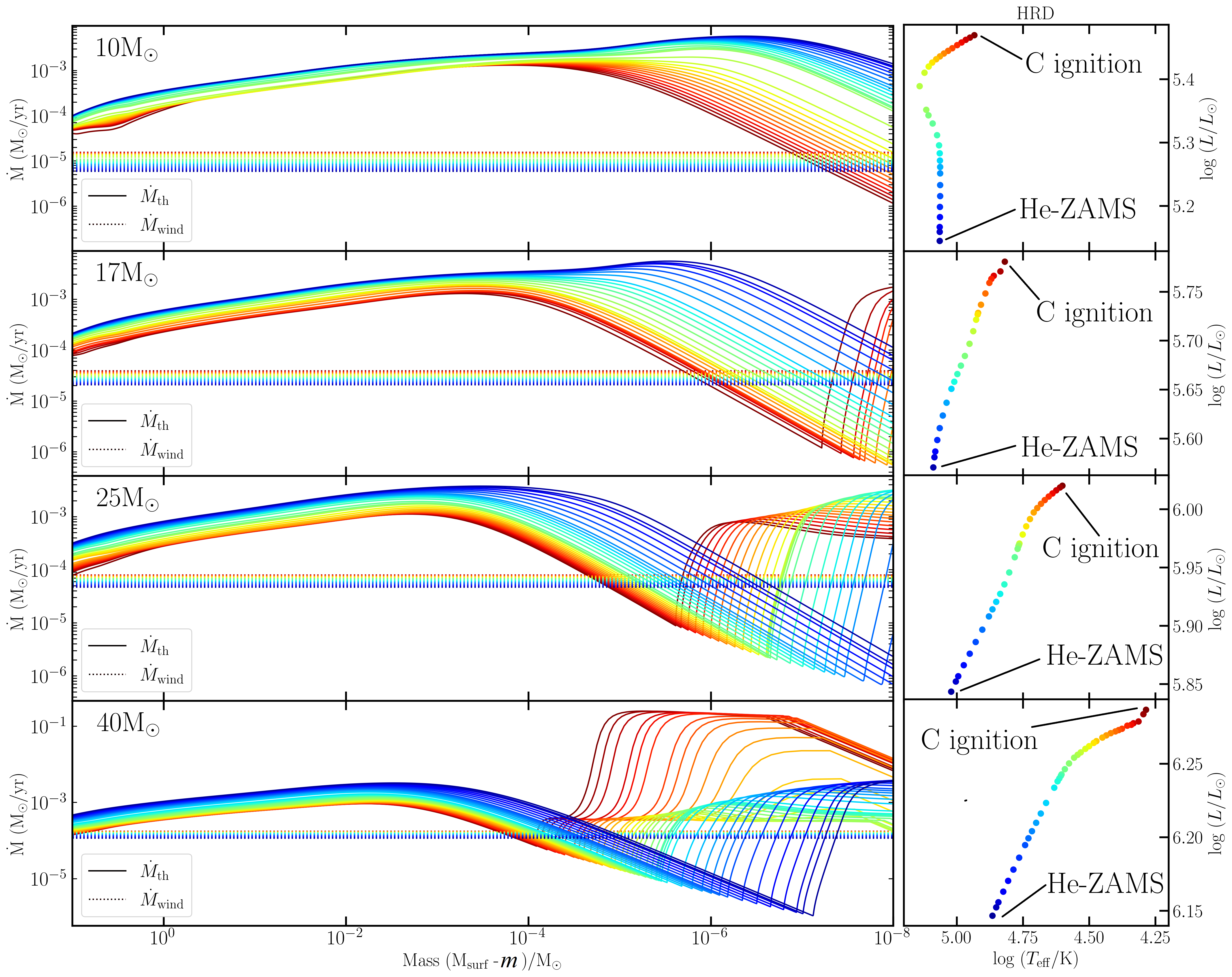}
	\caption{The structure of $\dot{M} _\textrm{th,crit}$ for $10\,M_{\odot}$, $17\,M_{\odot}$, $25\,M_{\odot}$ and $40\,M_{\odot}$ helium stars with no wind. The horizontal dot lines are the mass loss rates of $\eta=1$ Nugis \& Lamers wind. The colors of the different lines correspond to a specific helium-star model on the HRD on the right.
		\label{structure_10_17_25_40}}
\end{figure*}

\begin{figure*}[ht!]
	\plotone{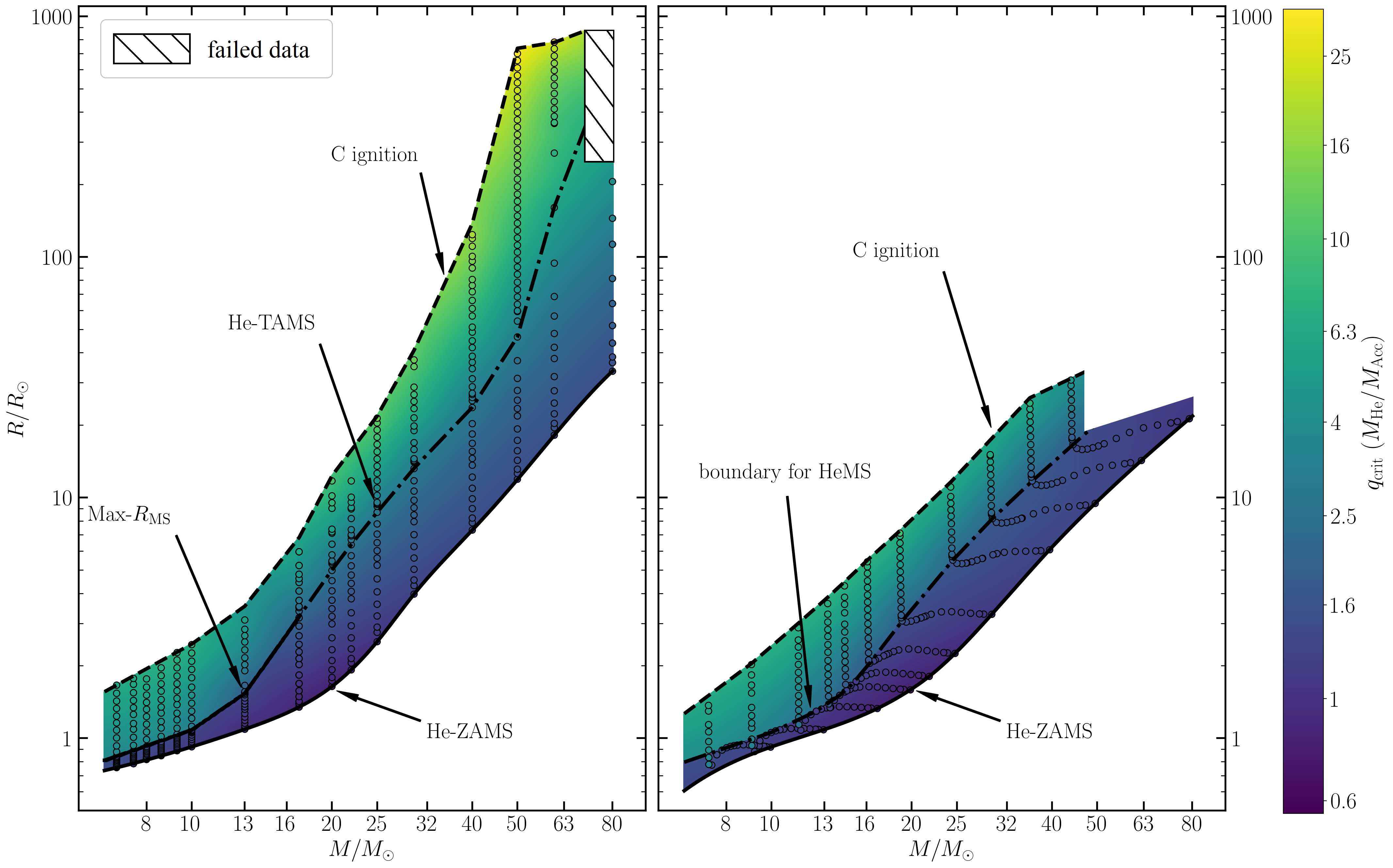}
	\caption{The $q_\textrm{crit}$ of specific stellar models and the RBF interpolate. The color represents the $q_\textrm{crit}$ and the text has a similar meaning to Figure\,\ref{MR-space}.
		\label{RBF-models}}
\end{figure*}

\section{The local thermal timescale profiles and the significance of wind decortication}

To show the significance of \textit{Wind-driven Decortication Effect} for massive helium stars with different masses and evolutionary stages, we give the comparison between $\dot{M} _\textrm{th,crit}$ for no wind helium stars and $\dot{M} _\textrm{wind}$ of $\eta=1$ Nugis \& Lamers wind. The panels of Figure\,\ref{structure_10_17_25_40} are the $\dot{M} _\textrm{th,crit}$ structures for different masses and evolution stages.

Based on the description of wind decortication in Section\,\ref{wind compression}, once the $\dot{M} _\textrm{wind}$ is lower than $\dot{M} _\textrm{th,crit}$, the wind shall force the surface to shrink until a new balance is built. In Figure\,\ref{structure_10_17_25_40} we find that with the increase of mass, the deeper that $\dot{M} _\textrm{wind}$ invades the surface layer of $\dot{M} _\textrm{th,crit}$ profile. The \textit{Wind-driven Decortication Effect} should be significant for massive stars and late-evolutionary stage stars. For $10\,M_{\odot}$ HeMS stars, wind decortication barely takes effect.

\section{RBF interpolation on M-R parameter space}

In this work, we use radial basis function (RBF) interpolation from \href{https://docs.scipy.org/doc/scipy/reference/generated/scipy.interpolate.Rbf.html}{\textbf{Scipy}} to calculate the $q_\textrm{crit}$ on $M-R$ parameter space and shift those interpolated results into $P_\textrm{orb}-q$ space. 

For no-wind models, we limited the HeMS space from the He-ZAMS to the model with the maximum HeMS radius. AS a result, if the stellar mass is lower than $17\,M_{\odot}$, some of the stellar models that are going through the shrinkage stage shall be folded and not participate in the interpolation. For the $\eta=0.8$ wind prescription, we artificially set an outer boundary space for all HeMS stars, and all models could be used as data for the interpolation. For HeHG models, the interpolation space of both wind prescriptions includes models from He-TAMS to the carbon ignition. For some late-evolutionary stage $80\,M_{\odot}$ models, due to a rapid shrinkage near the stellar surface during adiabatic mass loss, we could not calculate the $q_\textrm{crit}$ for them. We labeled them with the failed data.

From Figure\,\ref{RBF-models}, we have given the criterion for specific models from the massive helium star sequences and the results of the RBF interpolation. With the RBF smoothing parameter of 0.01, the errors of interpolation are quite low. Due to the minimum mass of the $\eta=0.8$ wind model being around $7\,M_{\odot}$, we add some intermediate-mass helium stars from Paper IV \citep{PaperIV} into the interpolation to make the parameter space comparable.

\section{Estimation of the non-conserved polytropic criterion}

For polytropic models, the $q_\textrm{crit}$ is determined by the stellar structure. The elemental composition does not take part in the calculation. 
Based on different mixing types of the envelope, MS stars are separated into two types: fully convective low mass stars ($<0.7\,M_{\odot}$) and radiative envelope massive stars ($>0.7\,M_{\odot}$). HeMS stars share a similar structure with massive MS stars. Therefore, they share the same $q_\textrm{crit}=1.6$ \citep{2014A&A...563A..83C,2007A&A...467.1181D} in polytropic models. The relationship of the criterion of HeHG and HG stars is just the same. We find their polytropic $q_\textrm{crit}=4$ \citep{2002MNRAS.329..897H,1997MNRAS.291..732T}.
Meanwhile, $q_\textrm{crit}=3$ for HeMS and MS stars is also widely used in BSE population synthesis code \citep{2023A&A...669A..82L,2002MNRAS.329..897H}. Here we note these value by
$q^{\textrm{con}}_\textrm{HeMS,1}=1.6$;  $q^{\textrm{con}}_\textrm{HeMS,2}=3.0$ and  $q^{\textrm{con}}_\textrm{HeHG}=4.0$

All these $q_\textrm{crit}$ above are conserved mass-transfer results. We use $q^{\textrm{con}}$ to represent the conserved transfer in the following text. The calculation follows such equations \citep{1997MNRAS.291..732T}:
\begin{equation}
	\zeta^{\textrm{con}}_{L} \left( q \right) = 2.13\,q-1.67,
\end{equation}
\begin{equation}
	\zeta_{ad}=\zeta^{\textrm{con}}_{L} \left( q^{\textrm{con}} \right),
\end{equation}
Where $\zeta_{ad}$ and $\zeta_L$ are the radius-mass exponents \citep{1985ibs..book...39W} for adiabatic response and Roche-lobe radius, which is defined by:
\begin{equation}
	\zeta_\textrm{ad} \equiv \left ( \frac{\partial \ln R_1}{\partial \ln M_1}  \right )_\textrm{ad} ; 
	\zeta_\textrm{L} \equiv \left ( \frac{\partial \ln R_{L_1}}{\partial \ln M_1}  \right ), 
\end{equation}
In this equation, $M_1$ is the mass of the donor, $R_1$ is the radius of the donor and $R_{L1}$ is the radius of the donor's Roche-lobe. $\zeta_L$ is influenced by the orbit angular momentum loss of the binary system, but $\zeta_{ad}$ is an adiabatic mechanism which depends on the stellar itself. We assume $\zeta_{ad}$ is isolated from the angular momentum loss. Here we use the method mentioned in \cite{2024A&A...681A..31P} to calculate the $\zeta_L$ when $\beta=1$:
\begin{equation}
	\zeta_{ad}=\zeta^{\beta=1}_{L}\left( q \right)=\zeta^{\textrm{con}}_{L}\left( q \right)-\frac{q}{q+1},
\end{equation}
\begin{equation}
	\zeta_{ad}=\zeta^{\beta=1}_{L} \left( q^{\beta=1} \right),
\end{equation}
Finally, we have the results of $\beta=1$ prescription in Figure\,\ref{Pq-space}: $q^{\beta=1}_\textrm{HeMS,1}=1.90$;  $q^{\beta=1}_\textrm{HeMS,2}=3.36$ and  $q^{\beta=1}_\textrm{HeHG}=4.38$. 

It should be noted that many BPS codes haven't considered such impacts of the specific angular momentum loss. Using the conserved criterion is an effective approximate method.


\bibliography{sample701}{}

@ARTICLE{PaperI,
	author = {{Ge}, Hongwei and {Hjellming}, Michael S. and {Webbink}, Ronald F. and {Chen}, Xuefei and {Han}, Zhanwen},
	title = "{Adiabatic Mass Loss in Binary Stars. I. Computational Method}",
	journal = {\apj},
	year = 2010,
	month = jul,
	volume = {717},
	number = {2},
	pages = {724-738},
	doi = {10.1088/0004-637X/717/2/724},
	archivePrefix = {arXiv},
	eprint = {1005.3099},
	primaryClass = {astro-ph.SR},
	adsurl = {https://ui.adsabs.harvard.edu/abs/2010ApJ...717..724G}
}

@ARTICLE{PaperII,
	author = {{Ge}, Hongwei and {Webbink}, Ronald F. and {Chen}, Xuefei and {Han}, Zhanwen},
	title = "{Adiabatic Mass Loss in Binary Stars. II. From Zero-age Main Sequence to the Base of the Giant Branch}",
	journal = {\apj},
	year = 2015,
	month = oct,
	volume = {812},
	number = {1},
	eid = {40},
	pages = {40},
	doi = {10.1088/0004-637X/812/1/40},
	archivePrefix = {arXiv},
	eprint = {1507.04843},
	primaryClass = {astro-ph.SR},
	adsurl = {https://ui.adsabs.harvard.edu/abs/2015ApJ...812...40G}
}

@ARTICLE{PaperIII,
	author = {{Ge}, Hongwei and {Webbink}, Ronald F. and {Chen}, Xuefei and {Han}, Zhanwen},
	title = "{Adiabatic Mass Loss in Binary Stars. III. From the Base of the Red Giant Branch to the Tip of the Asymptotic Giant Branch}",
	journal = {\apj},
	year = 2020,
	month = aug,
	volume = {899},
	number = {2},
	eid = {132},
	pages = {132},
	doi = {10.3847/1538-4357/aba7b7},
	archivePrefix = {arXiv},
	eprint = {2007.09848},
	primaryClass = {astro-ph.SR},
	adsurl = {https://ui.adsabs.harvard.edu/abs/2020ApJ...899..132G}
}

@ARTICLE{PaperIV,
	author = {{Zhang}, Lifu and {Ge}, Hongwei and {Chen}, Xuefei and {Han}, Zhanwen},
	title = "{Adiabatic Mass Loss in Binary Stars. IV. Low- and Intermediate-mass Helium Binary Stars}",
	journal = {\apjs},
	year = 2024,
	month = sep,
	volume = {274},
	number = {1},
	eid = {11},
	pages = {11},
	doi = {10.3847/1538-4365/ad6263},
	archivePrefix = {arXiv},
	eprint = {2406.13146},
	primaryClass = {astro-ph.SR},
	adsurl = {https://ui.adsabs.harvard.edu/abs/2024ApJS..274...11Z}
}

@ARTICLE{PaperV,
	author = {{Ge}, Hongwei and {Tout}, Christopher A. and {Chen}, Xuefei and {Wang}, Song and {Xiong}, Jianping and {Zhang}, Lifu and {Li}, Zhenwei and {Liu}, Qingzhong and {Han}, Zhanwen},
	title = "{Adiabatic Mass Loss in Binary Stars. V. Effects of Metallicity and Nonconservative Mass Transfer{\textemdash}Application in High Mass X-Ray Binaries}",
	journal = {\apj},
	year = 2024,
	month = nov,
	volume = {975},
	number = {2},
	eid = {254},
	pages = {254},
	doi = {10.3847/1538-4357/ad7ea6},
	archivePrefix = {arXiv},
	eprint = {2408.16350},
	primaryClass = {astro-ph.SR},
	adsurl = {https://ui.adsabs.harvard.edu/abs/2024ApJ...975..254G}
}

@ARTICLE{2023ApJ...945....7G,
       author = {{Ge}, Hongwei and {Tout}, Christopher A. and {Chen}, Xuefei and {Sarkar}, Arnab and {Walton}, Dominic J. and {Han}, Zhanwen},
        title = "{Criteria for Dynamical Timescale Mass Transfer of Metal-poor Intermediate-mass Stars}",
      journal = {\apj},
         year = 2023,
        month = mar,
       volume = {945},
       number = {1},
          eid = {7},
        pages = {7},
          doi = {10.3847/1538-4357/acb7e9},
archivePrefix = {arXiv},
       eprint = {2302.00183},
 primaryClass = {astro-ph.SR},
       adsurl = {https://ui.adsabs.harvard.edu/abs/2023ApJ...945....7G}
}

@ARTICLE{2012Sci...337..444S,
	author = {{Sana}, H. and {de Mink}, S.~E. and {de Koter}, A. and {Langer}, N. and {Evans}, C.~J. and {Gieles}, M. and {Gosset}, E. and {Izzard}, R.~G. and {Le Bouquin}, J. -B. and {Schneider}, F.~R.~N.},
	title = "{Binary Interaction Dominates the Evolution of Massive Stars}",
	journal = {Science},
	year = 2012,
	month = jul,
	volume = {337},
	number = {6093},
	pages = {444},
	doi = {10.1126/science.1223344},
	archivePrefix = {arXiv},
	eprint = {1207.6397},
	primaryClass = {astro-ph.SR},
	adsurl = {https://ui.adsabs.harvard.edu/abs/2012Sci...337..444S}
}

@ARTICLE{2017ApJS..230...15M,
	author = {{Moe}, Maxwell and {Di Stefano}, Rosanne},
	title = "{Mind Your Ps and Qs: The Interrelation between Period (P) and Mass-ratio (Q) Distributions of Binary Stars}",
	journal = {\apjs},
	year = 2017,
	month = jun,
	volume = {230},
	number = {2},
	eid = {15},
	pages = {15},
	doi = {10.3847/1538-4365/aa6fb6},
	archivePrefix = {arXiv},
	eprint = {1606.05347},
	primaryClass = {astro-ph.SR},
	adsurl = {https://ui.adsabs.harvard.edu/abs/2017ApJS..230...15M}
}

@ARTICLE{2024PrPNP.13404083C,
	author = {{Chen}， Xuefei and {Liu}, Zhengwei and {Han}, Zhanwen},
	title = "{Binary stars in the new millennium}",
	journal = {Progress in Particle and Nuclear Physics},
	year = 2024,
	month = jan,
	volume = {134},
	eid = {104083},
	pages = {104083},
	doi = {10.1016/j.ppnp.2023.104083},
	archivePrefix = {arXiv},
	eprint = {2311.11454},
	primaryClass = {astro-ph.SR},
	adsurl = {https://ui.adsabs.harvard.edu/abs/2024PrPNP.13404083C}
}

@ARTICLE{2001NewAR..45..135V,
	author = {{van der Hucht}, Karel A.},
	title = "{The VIIth catalogue of galactic Wolf-Rayet stars}",
	journal = {\nar},
	year = 2001,
	month = feb,
	volume = {45},
	number = {3},
	pages = {135-232},
	doi = {10.1016/S1387-6473(00)00112-3},
	adsurl = {https://ui.adsabs.harvard.edu/abs/2001NewAR..45..135V}
}

@ARTICLE{2015MNRAS.452.1068C,
       author = {{Chen}, Yang and {Bressan}, Alessandro and {Girardi}, L{\'e}o and {Marigo}, Paola and {Kong}, Xu and {Lanza}, Antonio},
        title = "{PARSEC evolutionary tracks of massive stars up to 350 M$_{{\ensuremath{\odot}}}$ at metallicities 0.0001 {\ensuremath{\leq}} Z {\ensuremath{\leq}} 0.04}",
      journal = {\mnras},
         year = 2015,
        month = sep,
       volume = {452},
       number = {1},
        pages = {1068-1080},
          doi = {10.1093/mnras/stv1281},
archivePrefix = {arXiv},
       eprint = {1506.01681},
 primaryClass = {astro-ph.SR},
       adsurl = {https://ui.adsabs.harvard.edu/abs/2015MNRAS.452.1068C}
}

@ARTICLE{2021MNRAS.503.4208S,
       author = {{Scott}, L.~J.~A. and {Hirschi}, R. and {Georgy}, C. and {Arnett}, W.~D. and {Meakin}, C. and {Kaiser}, E.~A. and {Ekstr{\"o}m}, S. and {Yusof}, N.},
        title = "{Convective core entrainment in 1D main-sequence stellar models}",
      journal = {\mnras},
         year = 2021,
        month = may,
       volume = {503},
       number = {3},
        pages = {4208-4220},
          doi = {10.1093/mnras/stab752},
archivePrefix = {arXiv},
       eprint = {2103.06196},
 primaryClass = {astro-ph.SR},
       adsurl = {https://ui.adsabs.harvard.edu/abs/2021MNRAS.503.4208S}
}

@ARTICLE{2019ApJ...870...77L,
       author = {{Li}, Yan and {Chen}, Xing-hao and {Chen}, Hai-liang},
        title = "{Overshooting in the Core Helium-burning Stage of a 30 M $_{{\ensuremath{\odot}}}$ Star Using the k-{\ensuremath{\omega}} Model}",
      journal = {\apj},
         year = 2019,
        month = jan,
       volume = {870},
       number = {2},
          eid = {77},
        pages = {77},
          doi = {10.3847/1538-4357/aaf1a5},
archivePrefix = {arXiv},
       eprint = {1811.06638},
 primaryClass = {astro-ph.SR},
       adsurl = {https://ui.adsabs.harvard.edu/abs/2019ApJ...870...77L}
}

@ARTICLE{2023ApJS..268...51L,
       author = {{Li}, Zhi and {Li}, Yan},
        title = "{Grids of Wolf-Rayet Stars Using MESA with the k - {\ensuremath{\omega}} Model: From 25 to 120 M $_{{\ensuremath{\odot}}}$ at Z = 0.02}",
      journal = {\apjs},
         year = 2023,
        month = oct,
       volume = {268},
       number = {2},
          eid = {51},
        pages = {51},
          doi = {10.3847/1538-4365/aced88},
       adsurl = {https://ui.adsabs.harvard.edu/abs/2023ApJS..268...51L}
}

@ARTICLE{2012A&A...537A.146E,
       author = {{Ekstr{\"o}m}, S. and {Georgy}, C. and {Eggenberger}, P. and {Meynet}, G. and {Mowlavi}, N. and {Wyttenbach}, A. and {Granada}, A. and {Decressin}, T. and {Hirschi}, R. and {Frischknecht}, U. and {Charbonnel}, C. and {Maeder}, A.},
        title = "{Grids of stellar models with rotation. I. Models from 0.8 to 120 M$_{{\ensuremath{\odot}}}$ at solar metallicity (Z = 0.014)}",
      journal = {\aap},
         year = 2012,
        month = jan,
       volume = {537},
          eid = {A146},
        pages = {A146},
          doi = {10.1051/0004-6361/201117751},
archivePrefix = {arXiv},
       eprint = {1110.5049},
 primaryClass = {astro-ph.SR},
       adsurl = {https://ui.adsabs.harvard.edu/abs/2012A&A...537A.146E}
}

@ARTICLE{2003A&A...404..975M,
       author = {{Meynet}, G. and {Maeder}, A.},
        title = "{Stellar evolution with rotation. X. Wolf-Rayet star populations at solar metallicity}",
      journal = {\aap},
         year = 2003,
        month = jun,
       volume = {404},
        pages = {975-990},
          doi = {10.1051/0004-6361:20030512},
archivePrefix = {arXiv},
       eprint = {astro-ph/0304069},
 primaryClass = {astro-ph},
       adsurl = {https://ui.adsabs.harvard.edu/abs/2003A&A...404..975M}
}

@INPROCEEDINGS{2022IAUS..366...21S,
       author = {{Sander}, Andreas A.~C. and {Vink}, Jorick S. and {Higgins}, Erin R. and {Shenar}, Tomer and {Hamann}, Wolf-Rainer and {Todt}, Helge},
        title = "{The origin and impact of Wolf-Rayet-type mass loss}",
    booktitle = {The Origin of Outflows in Evolved Stars},
         year = 2022,
       editor = {{Decin}, Leen and {Zijlstra}, Albert and {Gielen}, Clio},
       series = {IAU Symposium},
       volume = {366},
        month = jan,
        pages = {21-26},
          doi = {10.1017/S1743921322000400},
archivePrefix = {arXiv},
       eprint = {2202.04671},
 primaryClass = {astro-ph.SR},
       adsurl = {https://ui.adsabs.harvard.edu/abs/2022IAUS..366...21S}
}

@ARTICLE{2007ARA&A..45..177C,
       author = {{Crowther}, Paul A.},
        title = "{Physical Properties of Wolf-Rayet Stars}",
      journal = {\araa},
         year = 2007,
        month = sep,
       volume = {45},
       number = {1},
        pages = {177-219},
          doi = {10.1146/annurev.astro.45.051806.110615},
archivePrefix = {arXiv},
       eprint = {astro-ph/0610356},
 primaryClass = {astro-ph},
       adsurl = {https://ui.adsabs.harvard.edu/abs/2007ARA&A..45..177C}
}

@ARTICLE{1867CRAS...65..292W,
       author = {{Wolf}, C.~J.~E. and {Rayet}, G.},
        title = "{Spectroscopie stellaire}",
      journal = {Academie des Sciences Paris Comptes Rendus},
         year = 1867,
        month = jan,
       volume = {65},
        pages = {292-296},
       adsurl = {https://ui.adsabs.harvard.edu/abs/1867CRAS...65..292W}
}

@ARTICLE{1994A&A...287..803M,
	author = {{Maeder}, A. and {Meynet}, G.},
	title = "{New models of Wolf-Rayet stars and comparison with data in galaxies}",
	journal = {\aap},
	year = 1994,
	month = jul,
	volume = {287},
	pages = {803-816},
	adsurl = {https://ui.adsabs.harvard.edu/abs/1994A&A...287..803M}
}

@ARTICLE{2008A&A...485..245G,
       author = {{Groh}, J.~H. and {Oliveira}, A.~S. and {Steiner}, J.~E.},
        title = "{The qWR star HD 45166. II. Fundamental stellar parameters and evidence of a latitude-dependent wind}",
      journal = {\aap},
         year = 2008,
        month = jul,
       volume = {485},
       number = {1},
        pages = {245-256},
          doi = {10.1051/0004-6361:200809511},
archivePrefix = {arXiv},
       eprint = {0804.1766},
 primaryClass = {astro-ph},
       adsurl = {https://ui.adsabs.harvard.edu/abs/2008A&A...485..245G}
}

@ARTICLE{2001MNRAS.324...18B,
	author = {{Bartzakos}, P. and {Moffat}, A.~F.~J. and {Niemela}, V.~S.},
	title = "{Magellanic Cloud WC/WO Wolf-Rayet stars - I. Binary frequency and Roche lobe overflow formation}",
	journal = {\mnras},
	year = 2001,
	month = jun,
	volume = {324},
	number = {1},
	pages = {18-32},
	doi = {10.1046/j.1365-8711.2001.04126.x},
	adsurl = {https://ui.adsabs.harvard.edu/abs/2001MNRAS.324...18B}
}

@ARTICLE{2024ApJ...969..160L,
       author = {{Li}, Zhuowen and {Zhu}, Chunhua and {L{\"u}}, Guoliang and {Li}, Lin and {Liu}, Helei and {Guo}, Sufen and {Yu}, Jinlong and {Lu}, Xizhen},
        title = "{The Population Synthesis of Wolf{\textendash}Rayet Stars Involving Binary Merger Channels}",
      journal = {\apj},
         year = 2024,
        month = jul,
       volume = {969},
       number = {2},
          eid = {160},
        pages = {160},
          doi = {10.3847/1538-4357/ad4da8},
archivePrefix = {arXiv},
       eprint = {2405.11571},
 primaryClass = {astro-ph.SR},
       adsurl = {https://ui.adsabs.harvard.edu/abs/2024ApJ...969..160L}
}

@ARTICLE{2009ARA&A..47...63S,
       author = {{Smartt}, Stephen J.},
        title = "{Progenitors of Core-Collapse Supernovae}",
      journal = {\araa},
         year = 2009,
        month = sep,
       volume = {47},
       number = {1},
        pages = {63-106},
          doi = {10.1146/annurev-astro-082708-101737},
archivePrefix = {arXiv},
       eprint = {0908.0700},
 primaryClass = {astro-ph.SR},
       adsurl = {https://ui.adsabs.harvard.edu/abs/2009ARA&A..47...63S}
}

@ARTICLE{2015MNRAS.451.2123T,
       author = {{Tauris}, Thomas M. and {Langer}, Norbert and {Podsiadlowski}, Philipp},
        title = "{Ultra-stripped supernovae: progenitors and fate}",
      journal = {\mnras},
         year = 2015,
        month = aug,
       volume = {451},
       number = {2},
        pages = {2123-2144},
          doi = {10.1093/mnras/stv990},
archivePrefix = {arXiv},
       eprint = {1505.00270},
 primaryClass = {astro-ph.SR},
       adsurl = {https://ui.adsabs.harvard.edu/abs/2015MNRAS.451.2123T}
}

@ARTICLE{2018MNRAS.481.1908K,
       author = {{Kruckow}, Matthias U. and {Tauris}, Thomas M. and {Langer}, Norbert and {Kramer}, Michael and {Izzard}, Robert G.},
        title = "{Progenitors of gravitational wave mergers: binary evolution with the stellar grid-based code COMBINE}",
      journal = {\mnras},
         year = 2018,
        month = dec,
       volume = {481},
       number = {2},
        pages = {1908-1949},
          doi = {10.1093/mnras/sty2190},
archivePrefix = {arXiv},
       eprint = {1801.05433},
 primaryClass = {astro-ph.SR},
       adsurl = {https://ui.adsabs.harvard.edu/abs/2018MNRAS.481.1908K}
}

@ARTICLE{2016RAA....16..141Y,
       author = {{Yan}, Jing-Zhi and {Zhu}, Chun-Hua and {Wang}, Zhao-Jun and {L{\"u}}, Guo-Liang},
        title = "{The effects of convective overshooting on naked helium stars}",
      journal = {Research in Astronomy and Astrophysics},
         year = 2016,
        month = sep,
       volume = {16},
       number = {9},
          eid = {141},
        pages = {141},
          doi = {10.1088/1674-4527/16/9/141},
       adsurl = {https://ui.adsabs.harvard.edu/abs/2016RAA....16..141Y}
}

@ARTICLE{1988A&AS...72..259D,
       author = {{de Jager}, C. and {Nieuwenhuijzen}, H. and {van der Hucht}, K.~A.},
        title = "{Mass loss rates in the Hertzsprung-Russell diagram.}",
      journal = {\aaps},
         year = 1988,
        month = feb,
       volume = {72},
        pages = {259-289},
       adsurl = {https://ui.adsabs.harvard.edu/abs/1988A&AS...72..259D}
}

@ARTICLE{1995A&A...299..151H,
       author = {{Hamann}, W.-R. and {Koesterke}, L. and {Wessolowski}, U.},
        title = "{Spectral analyses of the Galactic Wolf-Rayet stars: hydrogen-helium abundances and improved stellar parameters for the WN class}",
      journal = {\aap},
         year = 1995,
        month = jul,
       volume = {299},
        pages = {151},
       adsurl = {https://ui.adsabs.harvard.edu/abs/1995A&A...299..151H}
}

@ARTICLE{2017A&A...607L...8V,
       author = {{Vink}, Jorick S.},
        title = "{Winds from stripped low-mass helium stars and Wolf-Rayet stars}",
      journal = {\aap},
         year = 2017,
        month = nov,
       volume = {607},
          eid = {L8},
        pages = {L8},
          doi = {10.1051/0004-6361/201731902},
archivePrefix = {arXiv},
       eprint = {1710.02010},
 primaryClass = {astro-ph.SR},
       adsurl = {https://ui.adsabs.harvard.edu/abs/2017A&A...607L...8V}
}

@ARTICLE{2017MNRAS.470.3970Y,
       author = {{Yoon}, Sung-Chul},
        title = "{Towards a better understanding of the evolution of Wolf-Rayet stars and Type Ib/Ic supernova progenitors}",
      journal = {\mnras},
         year = 2017,
        month = oct,
       volume = {470},
       number = {4},
        pages = {3970-3980},
          doi = {10.1093/mnras/stx1496},
archivePrefix = {arXiv},
       eprint = {1706.04716},
 primaryClass = {astro-ph.SR},
       adsurl = {https://ui.adsabs.harvard.edu/abs/2017MNRAS.470.3970Y}
}

@ARTICLE{2020MNRAS.499..873S,
       author = {{Sander}, Andreas A.~C. and {Vink}, Jorick S.},
        title = "{On the nature of massive helium star winds and Wolf-Rayet-type mass-loss}",
      journal = {\mnras},
         year = 2020,
        month = nov,
       volume = {499},
       number = {1},
        pages = {873-892},
          doi = {10.1093/mnras/staa2712},
archivePrefix = {arXiv},
       eprint = {2009.01849},
 primaryClass = {astro-ph.SR},
       adsurl = {https://ui.adsabs.harvard.edu/abs/2020MNRAS.499..873S}
}

@ARTICLE{2000A&A...360..227N,
	author = {{Nugis}, T. and {Lamers}, H.~J.~G.~L.~M.},
	title = "{Mass-loss rates of Wolf-Rayet stars as a function of stellar parameters}",
	journal = {\aap},
	year = 2000,
	month = aug,
	volume = {360},
	pages = {227-244},
	adsurl = {https://ui.adsabs.harvard.edu/abs/2000A&A...360..227N}
}

@BOOK{2023pbse.book.....T,
       author = {{Tauris}, Thomas M. and {van den Heuvel}, Edward P.~J.},
        title = "{Physics of Binary Star Evolution. From Stars to X-ray Binaries and Gravitational Wave Sources}",
         year = 2023,
          doi = {10.48550/arXiv.2305.09388},
       adsurl = {https://ui.adsabs.harvard.edu/abs/2023pbse.book.....T}
}

@INPROCEEDINGS{1976IAUS...73...75P,
	author = {{Paczynski}, B.},
	title = "{Common Envelope Binaries}",
	booktitle = {Structure and Evolution of Close Binary Systems},
	year = 1976,
	editor = {{Eggleton}, Peter and {Mitton}, Simon and {Whelan}, John},
	series = {IAU Symposium},
	volume = {73},
	month = jan,
	pages = {75},
	adsurl = {https://ui.adsabs.harvard.edu/abs/1976IAUS...73...75P}
}

@ARTICLE{1978A&A....62..317S,
	author = {{Savonije}, G.~J.},
	title = "{Roche-lobe overflow in X-ray binaries.}",
	journal = {\aap},
	year = 1978,
	month = jan,
	volume = {62},
	number = {3},
	pages = {317-338},
	adsurl = {https://ui.adsabs.harvard.edu/abs/1978A&A....62..317S}
}

@ARTICLE{1981ApJ...246..153M,
	author = {{Massey}, P.},
	title = "{The masses of Wolf-rayet stars.}",
	journal = {\apj},
	year = 1981,
	month = may,
	volume = {246},
	pages = {153-160},
	doi = {10.1086/158908},
	adsurl = {https://ui.adsabs.harvard.edu/abs/1981ApJ...246..153M}
}

@ARTICLE{1955ApJ...121..161S,
	author = {{Salpeter}, Edwin E.},
	title = "{The Luminosity Function and Stellar Evolution.}",
	journal = {\apj},
	year = 1955,
	month = jan,
	volume = {121},
	pages = {161},
	doi = {10.1086/145971},
	adsurl = {https://ui.adsabs.harvard.edu/abs/1955ApJ...121..161S}
}

@ARTICLE{2001MNRAS.322..231K,
	author = {{Kroupa}, Pavel},
	title = "{On the variation of the initial mass function}",
	journal = {\mnras},
	year = 2001,
	month = apr,
	volume = {322},
	number = {2},
	pages = {231-246},
	doi = {10.1046/j.1365-8711.2001.04022.x},
	archivePrefix = {arXiv},
	eprint = {astro-ph/0009005},
	primaryClass = {astro-ph},
	adsurl = {https://ui.adsabs.harvard.edu/abs/2001MNRAS.322..231K}
}

@ARTICLE{2020RAA....20..161H,
       author = {{Han}, Zhan-Wen and {Ge}, Hong-Wei and {Chen}, Xue-Fei and {Chen}, Hai-Liang},
        title = "{Binary Population Synthesis}",
      journal = {Research in Astronomy and Astrophysics},
         year = 2020,
        month = oct,
       volume = {20},
       number = {10},
          eid = {161},
        pages = {161},
          doi = {10.1088/1674-4527/20/10/161},
archivePrefix = {arXiv},
       eprint = {2009.08611},
 primaryClass = {astro-ph.SR},
       adsurl = {https://ui.adsabs.harvard.edu/abs/2020RAA....20..161H}
}

@ARTICLE{1987ApJ...318..794H,
	author = {{Hjellming}, Michael S. and {Webbink}, Ronald F.},
	title = "{Thresholds for Rapid Mass Transfer in Binary System. I. Polytropic Models}",
	journal = {\apj},
	year = 1987,
	month = jul,
	volume = {318},
	pages = {794},
	doi = {10.1086/165412},
	adsurl = {https://ui.adsabs.harvard.edu/abs/1987ApJ...318..794H}
}

@ARTICLE{1997A&A...327..620S,
	author = {{Soberman}, G.~E. and {Phinney}, E.~S. and {van den Heuvel}, E.~P.~J.},
	title = "{Stability criteria for mass transfer in binary stellar evolution.}",
	journal = {\aap},
	year = 1997,
	month = nov,
	volume = {327},
	pages = {620-635},
	doi = {10.48550/arXiv.astro-ph/9703016},
	archivePrefix = {arXiv},
	eprint = {astro-ph/9703016},
	primaryClass = {astro-ph},
	adsurl = {https://ui.adsabs.harvard.edu/abs/1997A&A...327..620S}
}

@INCOLLECTION{1985ibs..book...39W,
       author = {{Webbink}, R.~F.},
        title = "{Stellar evolution and binaries}",
    booktitle = {Interacting Binary Stars},
         year = 1985,
       editor = {{Pringle}, J.~E. and {Wade}, R.~A.},
        pages = {39},
       adsurl = {https://ui.adsabs.harvard.edu/abs/1985ibs..book...39W}
}

@ARTICLE{2002MNRAS.329..897H,
	author = {{Hurley}, Jarrod R. and {Tout}, Christopher A. and {Pols}, Onno R.},
	title = "{Evolution of binary stars and the effect of tides on binary populations}",
	journal = {\mnras},
	year = 2002,
	month = feb,
	volume = {329},
	number = {4},
	pages = {897-928},
	doi = {10.1046/j.1365-8711.2002.05038.x},
	archivePrefix = {arXiv},
	eprint = {astro-ph/0201220},
	primaryClass = {astro-ph},
	adsurl = {https://ui.adsabs.harvard.edu/abs/2002MNRAS.329..897H}
}

@ARTICLE{2014A&A...563A..83C,
	author = {{Claeys}, J.~S.~W. and {Pols}, O.~R. and {Izzard}, R.~G. and {Vink}, J. and {Verbunt}, F.~W.~M.},
	title = "{Theoretical uncertainties of the Type Ia supernova rate}",
	journal = {\aap},
	year = 2014,
	month = mar,
	volume = {563},
	eid = {A83},
	pages = {A83},
	doi = {10.1051/0004-6361/201322714},
	archivePrefix = {arXiv},
	eprint = {1401.2895},
	primaryClass = {astro-ph.SR},
	adsurl = {https://ui.adsabs.harvard.edu/abs/2014A&A...563A..83C}
}

@ARTICLE{2007A&A...467.1181D,
       author = {{de Mink}, S.~E. and {Pols}, O.~R. and {Hilditch}, R.~W.},
        title = "{Efficiency of mass transfer in massive close binaries. Tests from double-lined eclipsing binaries in the SMC}",
      journal = {\aap},
         year = 2007,
        month = jun,
       volume = {467},
       number = {3},
        pages = {1181-1196},
          doi = {10.1051/0004-6361:20067007},
archivePrefix = {arXiv},
       eprint = {astro-ph/0703480},
 primaryClass = {astro-ph},
       adsurl = {https://ui.adsabs.harvard.edu/abs/2007A&A...467.1181D}
}

@ARTICLE{2023A&A...669A..82L,
       author = {{Li}, Zhenwei and {Chen}, Xuefei and {Ge}, Hongwei and {Chen}, Hai-Liang and {Han}, Zhanwen},
        title = "{Influence of a mass transfer stability criterion on double white dwarf populations}",
      journal = {\aap},
         year = 2023,
        month = jan,
       volume = {669},
          eid = {A82},
        pages = {A82},
          doi = {10.1051/0004-6361/202243893},
archivePrefix = {arXiv},
       eprint = {2211.01861},
 primaryClass = {astro-ph.SR},
       adsurl = {https://ui.adsabs.harvard.edu/abs/2023A&A...669A..82L}
}

@ARTICLE{2024A&A...681A..31P,
       author = {{Picco}, A. and {Marchant}, P. and {Sana}, H. and {Nelemans}, G.},
        title = "{Forming merging double compact objects with stable mass transfer}",
      journal = {\aap},
         year = 2024,
        month = jan,
       volume = {681},
          eid = {A31},
        pages = {A31},
          doi = {10.1051/0004-6361/202347090},
archivePrefix = {arXiv},
       eprint = {2309.05736},
 primaryClass = {astro-ph.SR},
       adsurl = {https://ui.adsabs.harvard.edu/abs/2024A&A...681A..31P}
}

@ARTICLE{1971MNRAS.151..351E,
	author = {{Eggleton}, Peter P.},
	title = "{The evolution of low mass stars}",
	journal = {\mnras},
	year = 1971,
	month = jan,
	volume = {151},
	pages = {351},
	doi = {10.1093/mnras/151.3.351},
	adsurl = {https://ui.adsabs.harvard.edu/abs/1971MNRAS.151..351E}
}

@ARTICLE{1972MNRAS.156..361E,
	author = {{Eggleton}, Peter P.},
	title = "{Composition changes during stellar evolution}",
	journal = {\mnras},
	year = 1972,
	month = jan,
	volume = {156},
	pages = {361},
	doi = {10.1093/mnras/156.3.361},
	adsurl = {https://ui.adsabs.harvard.edu/abs/1972MNRAS.156..361E}
}

@ARTICLE{1973MNRAS.163..279E,
	author = {{Eggleton}, Peter P.},
	title = "{A numerical treatment of double shell source stars}",
	journal = {\mnras},
	year = 1973,
	month = jan,
	volume = {163},
	pages = {279},
	doi = {10.1093/mnras/163.3.279},
	adsurl = {https://ui.adsabs.harvard.edu/abs/1973MNRAS.163..279E}
}

@ARTICLE{2001A&A...373..555M,
	author = {{Maeder}, A. and {Meynet}, G.},
	title = "{Stellar evolution with rotation. VII. . Low metallicity models and the blue to red supergiant ratio in the SMC}",
	journal = {\aap},
	year = 2001,
	month = jul,
	volume = {373},
	pages = {555-571},
	doi = {10.1051/0004-6361:20010596},
	archivePrefix = {arXiv},
	eprint = {astro-ph/0105051},
	primaryClass = {astro-ph},
	adsurl = {https://ui.adsabs.harvard.edu/abs/2001A&A...373..555M}
}

@ARTICLE{1997MNRAS.285..696S,
	author = {{Schroder}, Klaus-Peter and {Pols}, Onno R. and {Eggleton}, Peter P.},
	title = "{A critical test of stellar evolution and convective core `overshooting' by means of zeta Aurigae systems}",
	journal = {\mnras},
	year = 1997,
	month = mar,
	volume = {285},
	number = {4},
	pages = {696-710},
	doi = {10.1093/mnras/285.4.696},
	adsurl = {https://ui.adsabs.harvard.edu/abs/1997MNRAS.285..696S}
}

@ARTICLE{1998MNRAS.298..525P,
	author = {{Pols}, Onno R. and {Schr{\"o}der}, Klaus-Peter and {Hurley}, Jarrod R. and {Tout}, Christopher A. and {Eggleton}, Peter P.},
	title = "{Stellar evolution models for Z = 0.0001 to 0.03}",
	journal = {\mnras},
	year = 1998,
	month = aug,
	volume = {298},
	number = {2},
	pages = {525-536},
	doi = {10.1046/j.1365-8711.1998.01658.x},
	adsurl = {https://ui.adsabs.harvard.edu/abs/1998MNRAS.298..525P}
}

@ARTICLE{2019ApJ...878...49W,
	author = {{Woosley}, S.~E.},
	title = "{The Evolution of Massive Helium Stars, Including Mass Loss}",
	journal = {\apj},
	year = 2019,
	month = jun,
	volume = {878},
	number = {1},
	eid = {49},
	pages = {49},
	doi = {10.3847/1538-4357/ab1b41},
	archivePrefix = {arXiv},
	eprint = {1901.00215},
	primaryClass = {astro-ph.SR},
	adsurl = {https://ui.adsabs.harvard.edu/abs/2019ApJ...878...49W}
}

@ARTICLE{1992ApJ...397..717I,
       author = {{Iglesias}, Carlos A. and {Rogers}, Forrest J. and {Wilson}, Brian G.},
        title = "{Spin-Orbit Interaction Effects on the Rosseland Mean Opacity}",
      journal = {\apj},
         year = 1992,
        month = oct,
       volume = {397},
        pages = {717},
          doi = {10.1086/171827},
       adsurl = {https://ui.adsabs.harvard.edu/abs/1992ApJ...397..717I}
}

@ARTICLE{2006A&A...450..219P,
       author = {{Petrovic}, J. and {Pols}, O. and {Langer}, N.},
        title = "{Are luminous and metal-rich Wolf-Rayet stars inflated?}",
      journal = {\aap},
         year = 2006,
        month = apr,
       volume = {450},
       number = {1},
        pages = {219-225},
          doi = {10.1051/0004-6361:20035837},
       adsurl = {https://ui.adsabs.harvard.edu/abs/2006A&A...450..219P}
}

@ARTICLE{2012A&A...538A..40G,
       author = {{Gr{\"a}fener}, G. and {Owocki}, S.~P. and {Vink}, J.~S.},
        title = "{Stellar envelope inflation near the Eddington limit. Implications for the radii of Wolf-Rayet stars and luminous blue variables}",
      journal = {\aap},
         year = 2012,
        month = feb,
       volume = {538},
          eid = {A40},
        pages = {A40},
          doi = {10.1051/0004-6361/201117497},
archivePrefix = {arXiv},
       eprint = {1112.1910},
 primaryClass = {astro-ph.SR},
       adsurl = {https://ui.adsabs.harvard.edu/abs/2012A&A...538A..40G}
}

@ARTICLE{2013ApJS..208....4P,
       author = {{Paxton}, Bill and {Cantiello}, Matteo and {Arras}, Phil and {Bildsten}, Lars and {Brown}, Edward F. and {Dotter}, Aaron and {Mankovich}, Christopher and {Montgomery}, M.~H. and {Stello}, Dennis and {Timmes}, F.~X. and {Townsend}, Richard},
        title = "{Modules for Experiments in Stellar Astrophysics (MESA): Planets, Oscillations, Rotation, and Massive Stars}",
      journal = {\apjs},
         year = 2013,
        month = sep,
       volume = {208},
       number = {1},
          eid = {4},
        pages = {4},
          doi = {10.1088/0067-0049/208/1/4},
archivePrefix = {arXiv},
       eprint = {1301.0319},
 primaryClass = {astro-ph.SR},
       adsurl = {https://ui.adsabs.harvard.edu/abs/2013ApJS..208....4P}
}

@ARTICLE{2016ApJ...821..109R,
       author = {{Ro}, Stephen and {Matzner}, Christopher D.},
        title = "{On the Launching and Structure of Radiatively Driven Winds in Wolf-Rayet Stars}",
      journal = {\apj},
         year = 2016,
        month = apr,
       volume = {821},
       number = {2},
          eid = {109},
        pages = {109},
          doi = {10.3847/0004-637X/821/2/109},
archivePrefix = {arXiv},
       eprint = {1602.07318},
 primaryClass = {astro-ph.SR},
       adsurl = {https://ui.adsabs.harvard.edu/abs/2016ApJ...821..109R}
}

@ARTICLE{2020ApJ...890...51E,
	author = {{Ertl}, T. and {Woosley}, S.~E. and {Sukhbold}, Tuguldur and {Janka}, H. -T.},
	title = "{The Explosion of Helium Stars Evolved with Mass Loss}",
	journal = {\apj},
	year = 2020,
	month = feb,
	volume = {890},
	number = {1},
	eid = {51},
	pages = {51},
	doi = {10.3847/1538-4357/ab6458},
	archivePrefix = {arXiv},
	eprint = {1910.01641},
	primaryClass = {astro-ph.HE},
	adsurl = {https://ui.adsabs.harvard.edu/abs/2020ApJ...890...51E}
}

@BOOK{2013sse..book.....K,
	author = {{Kippenhahn}, Rudolf and {Weigert}, Alfred and {Weiss}, Achim},
	title = "{Stellar Structure and Evolution}",
	year = 2013,
	doi = {10.1007/978-3-642-30304-3},
	adsurl = {https://ui.adsabs.harvard.edu/abs/2013sse..book.....K}
}

@ARTICLE{2023A&A...669A..45T,
	author = {{Temmink}, K.~D. and {Pols}, O.~R. and {Justham}, S. and {Istrate}, A.~G. and {Toonen}, S.},
	title = "{Coping with loss. Stability of mass transfer from post-main-sequence donor stars}",
	journal = {\aap},
	year = 2023,
	month = jan,
	volume = {669},
	eid = {A45},
	pages = {A45},
	doi = {10.1051/0004-6361/202244137},
	archivePrefix = {arXiv},
	eprint = {2209.12707},
	primaryClass = {astro-ph.SR},
	adsurl = {https://ui.adsabs.harvard.edu/abs/2023A&A...669A..45T}
}

@ARTICLE{1973A&A....25..387V,
       author = {{van den Heuvel}, E.~P.~J. and {De Loore}, C.},
        title = "{The nature of X-ray binaries III. Evolution of massive close binaries with one collapsed component - with a possible application to Cygnus X-3.}",
      journal = {\aap},
         year = 1973,
        month = jun,
       volume = {25},
        pages = {387},
       adsurl = {https://ui.adsabs.harvard.edu/abs/1973A&A....25..387V}
}

@ARTICLE{1991PhR...203....1B,
       author = {{Bhattacharya}, D. and {van den Heuvel}, E.~P.~J.},
        title = "{Formation and evolution of binary and millisecond radio pulsars}",
      journal = {\physrep},
         year = 1991,
        month = jan,
       volume = {203},
       number = {1-2},
        pages = {1-124},
          doi = {10.1016/0370-1573(91)90064-S},
       adsurl = {https://ui.adsabs.harvard.edu/abs/1991PhR...203....1B}
}

@ARTICLE{2023A&A...674A.216L,
       author = {{Lu}, Xizhen and {Zhu}, Chunhua and {Liu}, Helei and {Guo}, Sufen and {Yu}, Jinlong and {L{\"u}}, Guoliang},
        title = "{Hydrogen-free Wolf-Rayet stars: Helium stars with envelope-inflation structure and rotation}",
      journal = {\aap},
         year = 2023,
        month = jun,
       volume = {674},
          eid = {A216},
        pages = {A216},
          doi = {10.1051/0004-6361/202243188},
archivePrefix = {arXiv},
       eprint = {2304.05897},
 primaryClass = {astro-ph.SR},
       adsurl = {https://ui.adsabs.harvard.edu/abs/2023A&A...674A.216L}
}

@ARTICLE{2017ApJ...846..170T,
	author = {{Tauris}, T.~M. and {Kramer}, M. and {Freire}, P.~C.~C. and {Wex}, N. and {Janka}, H.-T. and {Langer}, N. and {Podsiadlowski}, Ph. and {Bozzo}, E. and {Chaty}, S. and {Kruckow}, M.~U. and {van den Heuvel}, E.~P.~J. and {Antoniadis}, J. and {Breton}, R.~P. and {Champion}, D.~J.},
	title = "{Formation of Double Neutron Star Systems}",
	journal = {\apj},
	year = 2017,
	month = sep,
	volume = {846},
	number = {2},
	eid = {170},
	pages = {170},
	doi = {10.3847/1538-4357/aa7e89},
	archivePrefix = {arXiv},
	eprint = {1706.09438},
	primaryClass = {astro-ph.HE},
	adsurl = {https://ui.adsabs.harvard.edu/abs/2017ApJ...846..170T}
}

@ARTICLE{2018MNRAS.481.4009V,
	author = {{Vigna-G{\'o}mez}, Alejandro and {Neijssel}, Coenraad J. and {Stevenson}, Simon and {Barrett}, Jim W. and {Belczynski}, Krzysztof and {Justham}, Stephen and {de Mink}, Selma E. and {M{\"u}ller}, Bernhard and {Podsiadlowski}, Philipp and {Renzo}, Mathieu and {Sz{\'e}csi}, Dorottya and {Mandel}, Ilya},
	title = "{On the formation history of Galactic double neutron stars}",
	journal = {\mnras},
	year = 2018,
	month = dec,
	volume = {481},
	number = {3},
	pages = {4009-4029},
	doi = {10.1093/mnras/sty2463},
	archivePrefix = {arXiv},
	eprint = {1805.07974},
	primaryClass = {astro-ph.SR},
	adsurl = {https://ui.adsabs.harvard.edu/abs/2018MNRAS.481.4009V}
}

@ARTICLE{2019MNRAS.490.3740N,
	author = {{Neijssel}, Coenraad J. and {Vigna-G{\'o}mez}, Alejandro and {Stevenson}, Simon and {Barrett}, Jim W. and {Gaebel}, Sebastian M. and {Broekgaarden}, Floor S. and {de Mink}, Selma E. and {Sz{\'e}csi}, Dorottya and {Vinciguerra}, Serena and {Mandel}, Ilya},
	title = "{The effect of the metallicity-specific star formation history on double compact object mergers}",
	journal = {\mnras},
	year = 2019,
	month = dec,
	volume = {490},
	number = {3},
	pages = {3740-3759},
	doi = {10.1093/mnras/stz2840},
	archivePrefix = {arXiv},
	eprint = {1906.08136},
	primaryClass = {astro-ph.SR},
	adsurl = {https://ui.adsabs.harvard.edu/abs/2019MNRAS.490.3740N}
}

@ARTICLE{1997MNRAS.291..732T,
       author = {{Tout}, Christopher A. and {Aarseth}, Sverre J. and {Pols}, Onno R. and {Eggleton}, Peter P.},
        title = "{Rapid binary star evolution for N-body simulations and population synthesis}",
      journal = {\mnras},
         year = 1997,
        month = nov,
       volume = {291},
       number = {4},
        pages = {732-748},
          doi = {10.1093/mnras/291.4.732},
       adsurl = {https://ui.adsabs.harvard.edu/abs/1997MNRAS.291..732T}
}

@ARTICLE{2019A&A...625A..57H,
	author = {{Hamann}, W.-R. and {Gr{\"a}fener}, G. and {Liermann}, A. and {Hainich}, R. and {Sander}, A.~A.~C. and {Shenar}, T. and {Ramachandran}, V. and {Todt}, H. and {Oskinova}, L.~M.},
	title = "{The Galactic WN stars revisited. Impact of Gaia distances on fundamental stellar parameters}",
	journal = {\aap},
	year = 2019,
	month = may,
	volume = {625},
	eid = {A57},
	pages = {A57},
	doi = {10.1051/0004-6361/201834850},
	archivePrefix = {arXiv},
	eprint = {1904.04687},
	primaryClass = {astro-ph.SR},
	adsurl = {https://ui.adsabs.harvard.edu/abs/2019A&A...625A..57H}
}

@ARTICLE{2012ARA&A..50..107L,
	author = {{Langer}, N.},
	title = "{Presupernova Evolution of Massive Single and Binary Stars}",
	journal = {\araa},
	year = 2012,
	month = sep,
	volume = {50},
	pages = {107-164},
	doi = {10.1146/annurev-astro-081811-125534},
	archivePrefix = {arXiv},
	eprint = {1206.5443},
	primaryClass = {astro-ph.SR},
	adsurl = {https://ui.adsabs.harvard.edu/abs/2012ARA&A..50..107L}
}

@ARTICLE{2011A&A...535A..56G,
	author = {{Gr{\"a}fener}, G. and {Vink}, J.~S. and {de Koter}, A. and {Langer}, N.},
	title = "{The Eddington factor as the key to understand the winds of the most massive stars. Evidence for a {\ensuremath{\Gamma}}-dependence of Wolf-Rayet type mass loss}",
	journal = {\aap},
	year = 2011,
	month = nov,
	volume = {535},
	eid = {A56},
	pages = {A56},
	doi = {10.1051/0004-6361/201116701},
	archivePrefix = {arXiv},
	eprint = {1106.5361},
	primaryClass = {astro-ph.SR},
	adsurl = {https://ui.adsabs.harvard.edu/abs/2011A&A...535A..56G}
}

@ARTICLE{2012MNRAS.424.1601F,
	author = {{Fahed}, R. and {Moffat}, A.~F.~J.},
	title = "{Colliding winds in five WR+O systems of the Southern hemisphere}",
	journal = {\mnras},
	year = 2012,
	month = aug,
	volume = {424},
	number = {3},
	pages = {1601-1613},
	doi = {10.1111/j.1365-2966.2012.20494.x},
	adsurl = {https://ui.adsabs.harvard.edu/abs/2012MNRAS.424.1601F}
}

@ARTICLE{1996A&A...306..771R,
	author = {{Rauw}, G. and {Vreux}, J.-M. and {Gosset}, E. and {Hutsemekers}, D. and {Magain}, P. and {Rochowicz}, K.},
	title = "{WR22: the most massive Wolf-Rayet star ever weighed.}",
	journal = {\aap},
	year = 1996,
	month = feb,
	volume = {306},
	pages = {771},
	adsurl = {https://ui.adsabs.harvard.edu/abs/1996A&A...306..771R}
}

@INPROCEEDINGS{2008RMxAC..33...91G,
	author = {{Gamen}, R. and {Gosset}, E. and {Morrell}, N.~I. and {Niemela}, V.~S. and {Sana}, H. and {Naz{\'e}}, Y. and {Rauw}, G. and {Barb{\'a}}, R.~H. and {Solivella}, G.~R.},
	title = "{The first orbital solution for the massive colliding-wind binary HD 93162 ({\ensuremath{\equiv}} WR 25)}",
	booktitle = {Revista Mexicana de Astronomia y Astrofisica Conference Series},
	year = 2008,
	series = {Revista Mexicana de Astronomia y Astrofisica Conference Series},
	volume = {33},
	month = aug,
	pages = {91-93},
	adsurl = {https://ui.adsabs.harvard.edu/abs/2008RMxAC..33...91G}
}

@ARTICLE{2017MNRAS.467.3105M,
	author = {{Munoz}, Melissa and {Moffat}, Anthony F.~J. and {Hill}, Grant M. and {Shenar}, Tomer and {Richardson}, Noel D. and {Pablo}, Herbert and {St-Louis}, Nicole and {Ramiaramanantsoa}, Tahina},
	title = "{WR 148: identifying the companion of an extreme runaway massive binary$^{*}$}",
	journal = {\mnras},
	year = 2017,
	month = may,
	volume = {467},
	number = {3},
	pages = {3105-3121},
	doi = {10.1093/mnras/stw2283},
	archivePrefix = {arXiv},
	eprint = {1609.08289},
}

@ARTICLE{2013A&A...552A..22C,
	author = {{Collado}, A. and {Gamen}, R. and {Barb{\'a}}, R.~H.},
	title = "{The new Wolf-Rayet binary system WR62a}",
	journal = {\aap},
	year = 2013,
	month = apr,
	volume = {552},
	eid = {A22},
	pages = {A22},
	doi = {10.1051/0004-6361/201118460},
	archivePrefix = {arXiv},
	eprint = {1303.2914},
	primaryClass = {astro-ph.SR},
	adsurl = {https://ui.adsabs.harvard.edu/abs/2013A&A...552A..22C}
}

@ARTICLE{1996A&A...314..521V,
	author = {{van Kerkwijk}, M.~H. and {Geballe}, T.~R. and {King}, D.~L. and {van der Klis}, M. and {van Paradijs}, J.},
	title = "{The Wolf-Rayet counterpart of Cygnus X-3.}",
	journal = {\aap},
	year = 1996,
	month = oct,
	volume = {314},
	pages = {521-540},
	doi = {10.48550/arXiv.astro-ph/9604100},
	archivePrefix = {arXiv},
	eprint = {astro-ph/9604100},
	primaryClass = {astro-ph},
	adsurl = {https://ui.adsabs.harvard.edu/abs/1996A&A...314..521V}
}

@ARTICLE{2009A&A...501..679V,
	author = {{Vilhu}, O. and {Hakala}, P. and {Hannikainen}, D.~C. and {McCollough}, M. and {Koljonen}, K.},
	title = "{Orbital modulation of X-ray emission lines in Cygnus X-3}",
	journal = {\aap},
	year = 2009,
	month = jul,
	volume = {501},
	number = {2},
	pages = {679-686},
	doi = {10.1051/0004-6361/200811293},
	archivePrefix = {arXiv},
	eprint = {0904.3967},
	primaryClass = {astro-ph.HE},
	adsurl = {https://ui.adsabs.harvard.edu/abs/2009A&A...501..679V}
}

@ARTICLE{2010ApJ...718..488S,
	author = {{Shrader}, Chris R. and {Titarchuk}, Lev and {Shaposhnikov}, Nikolai},
	title = "{New Evidence for a Black Hole in the Compact Binary Cygnus X-3}",
	journal = {\apj},
	year = 2010,
	month = jul,
	volume = {718},
	number = {1},
	pages = {488-493},
	doi = {10.1088/0004-637X/718/1/488},
	archivePrefix = {arXiv},
	eprint = {1005.5362},
	primaryClass = {astro-ph.HE},
	adsurl = {https://ui.adsabs.harvard.edu/abs/2010ApJ...718..488S}
}

@ARTICLE{2012MNRAS.426.1031Z,
	author = {{Zdziarski}, Andrzej A. and {Maitra}, Chandreyee and {Frankowski}, Adam and {Skinner}, Gerald K. and {Misra}, Ranjeev},
	title = "{Energy-dependent orbital modulation of X-rays and constraints on emission of the jet in Cyg X-3}",
	journal = {\mnras},
	year = 2012,
	month = oct,
	volume = {426},
	number = {2},
	pages = {1031-1042},
	doi = {10.1111/j.1365-2966.2012.21635.x},
	archivePrefix = {arXiv},
	eprint = {1205.4402},
	primaryClass = {astro-ph.HE},
	adsurl = {https://ui.adsabs.harvard.edu/abs/2012MNRAS.426.1031Z}
}

@ARTICLE{2007ApJ...669L..21P,
	author = {{Prestwich}, A.~H. and {Kilgard}, R. and {Crowther}, P.~A. and {Carpano}, S. and {Pollock}, A.~M.~T. and {Zezas}, A. and {Saar}, S.~H. and {Roberts}, T.~P. and {Ward}, M.~J.},
	title = "{The Orbital Period of the Wolf-Rayet Binary IC 10 X-1: Dynamic Evidence that the Compact Object Is a Black Hole}",
	journal = {\apjl},
	year = 2007,
	month = nov,
	volume = {669},
	number = {1},
	pages = {L21-L24},
	doi = {10.1086/523755},
	archivePrefix = {arXiv},
	eprint = {0709.2892},
	primaryClass = {astro-ph},
	adsurl = {https://ui.adsabs.harvard.edu/abs/2007ApJ...669L..21P}
}

@ARTICLE{2013Natur.503..500L,
	author = {{Liu}, Ji-Feng and {Bregman}, Joel N. and {Bai}, Yu and {Justham}, Stephen and {Crowther}, Paul},
	title = "{Puzzling accretion onto a black hole in the ultraluminous X-ray source M 101 ULX-1}",
	journal = {\nat},
	year = 2013,
	month = nov,
	volume = {503},
	number = {7477},
	pages = {500-503},
	doi = {10.1038/nature12762},
	archivePrefix = {arXiv},
	eprint = {1312.0337},
	primaryClass = {astro-ph.HE},
	adsurl = {https://ui.adsabs.harvard.edu/abs/2013Natur.503..500L}
}

@ARTICLE{2010MNRAS.403L..41C,
	author = {{Crowther}, P.~A. and {Barnard}, R. and {Carpano}, S. and {Clark}, J.~S. and {Dhillon}, V.~S. and {Pollock}, A.~M.~T.},
	title = "{NGC 300 X-1 is a Wolf-Rayet/black hole binary}",
	journal = {\mnras},
	year = 2010,
	month = mar,
	volume = {403},
	number = {1},
	pages = {L41-L45},
	doi = {10.1111/j.1745-3933.2010.00811.x},
	archivePrefix = {arXiv},
	eprint = {1001.4616},
	primaryClass = {astro-ph.SR},
	adsurl = {https://ui.adsabs.harvard.edu/abs/2010MNRAS.403L..41C}
}

@ARTICLE{2004ApJ...605..360W,
	author = {{Weisskopf}, Martin C. and {Wu}, Kinwah and {Tennant}, Allyn F. and {Swartz}, Douglas A. and {Ghosh}, Kajal K.},
	title = "{On the Nature of the Bright Short-Period X-Ray Source in the Circinus Galaxy Field}",
	journal = {\apj},
	year = 2004,
	month = apr,
	volume = {605},
	number = {1},
	pages = {360-367},
	doi = {10.1086/381307},
	archivePrefix = {arXiv},
	eprint = {astro-ph/0311291},
	primaryClass = {astro-ph},
	adsurl = {https://ui.adsabs.harvard.edu/abs/2004ApJ...605..360W}
}

@ARTICLE{2015MNRAS.452.1112E,
	author = {{Esposito}, P. and {Israel}, G.~L. and {Milisavljevic}, D. and {Mapelli}, M. and {Zampieri}, L. and {Sidoli}, L. and {Fabbiano}, G. and {Rodr{\'\i}guez Castillo}, G.~A.},
	title = "{Periodic signals from the Circinus region: two new cataclysmic variables and the ultraluminous X-ray source candidate GC X-1}",
	journal = {\mnras},
	year = 2015,
	month = sep,
	volume = {452},
	number = {2},
	pages = {1112-1127},
	doi = {10.1093/mnras/stv1379},
	archivePrefix = {arXiv},
	eprint = {1506.05808},
	primaryClass = {astro-ph.HE},
	adsurl = {https://ui.adsabs.harvard.edu/abs/2015MNRAS.452.1112E}
}

@ARTICLE{1989A&A...210...93L,
       author = {{Langer}, N.},
        title = "{Standard models of Wolf-Rayet stars.}",
      journal = {\aap},
         year = 1989,
        month = feb,
       volume = {210},
        pages = {93-113},
       adsurl = {https://ui.adsabs.harvard.edu/abs/1989A&A...210...93L}
}

@ARTICLE{2017MNRAS.471.4256V,
       author = {{van den Heuvel}, E.~P.~J. and {Portegies Zwart}, S.~F. and {de Mink}, S.~E.},
        title = "{Forming short-period Wolf-Rayet X-ray binaries and double black holes through stable mass transfer}",
      journal = {\mnras},
         year = 2017,
        month = nov,
       volume = {471},
       number = {4},
        pages = {4256-4264},
          doi = {10.1093/mnras/stx1430},
archivePrefix = {arXiv},
       eprint = {1701.02355},
 primaryClass = {astro-ph.SR},
       adsurl = {https://ui.adsabs.harvard.edu/abs/2017MNRAS.471.4256V}
}

@ARTICLE{2025ApJ...979..112N,
       author = {{Nie}, Yu-Dong and {Shao}, Yong and {He}, Jian-Guo and {Wei}, Ze-Lin and {Xu}, Xiao-Jie and {Li}, Xiang-Dong},
        title = "{Modeling High Mass X-Ray Binaries to Double Neutron Stars through Common Envelope Evolution}",
      journal = {\apj},
         year = 2025,
        month = feb,
       volume = {979},
       number = {2},
          eid = {112},
        pages = {112},
          doi = {10.3847/1538-4357/ad9a65},
archivePrefix = {arXiv},
       eprint = {2412.01776},
 primaryClass = {astro-ph.SR},
       adsurl = {https://ui.adsabs.harvard.edu/abs/2025ApJ...979..112N}
}

@ARTICLE{1879RSPS...29..168D,
       author = {{Darwin}, G.~H.},
        title = "{The Determination of the Secular Effects of Tidal Friction by a Graphical Method}",
      journal = {Proceedings of the Royal Society of London Series I},
         year = 1879,
        month = jan,
       volume = {29},
        pages = {168-181},
       adsurl = {https://ui.adsabs.harvard.edu/abs/1879RSPS...29..168D}
}

@ARTICLE{1995ApJ...444L..41R,
       author = {{Rasio}, Frederic A.},
        title = "{The Minimum Mass Ratio of W Ursae Majoris Binaries}",
      journal = {\apjl},
         year = 1995,
        month = may,
       volume = {444},
        pages = {L41},
          doi = {10.1086/187855},
archivePrefix = {arXiv},
       eprint = {astro-ph/9502028},
 primaryClass = {astro-ph},
       adsurl = {https://ui.adsabs.harvard.edu/abs/1995ApJ...444L..41R}
}

@ARTICLE{2001ApJ...562.1012E,
       author = {{Eggleton}, Peter P. and {Kiseleva-Eggleton}, Ludmila},
        title = "{Orbital Evolution in Binary and Triple Stars, with an Application to SS Lacertae}",
      journal = {\apj},
         year = 2001,
        month = dec,
       volume = {562},
       number = {2},
        pages = {1012-1030},
          doi = {10.1086/323843},
archivePrefix = {arXiv},
       eprint = {astro-ph/0104126},
 primaryClass = {astro-ph},
       adsurl = {https://ui.adsabs.harvard.edu/abs/2001ApJ...562.1012E}
}

@ARTICLE{2019arXiv190701877S,
       author = {{Sargsyan}, V.~V. and {Lenske}, H. and {Adamian}, G.~G. and {Antonenko}, N.~V.},
        title = "{On the Darwin instability effect in binary systems}",
      journal = {arXiv e-prints},
         year = 2019,
        month = jul,
          eid = {arXiv:1907.01877},
        pages = {arXiv:1907.01877},
          doi = {10.48550/arXiv.1907.01877},
archivePrefix = {arXiv},
       eprint = {1907.01877},
 primaryClass = {astro-ph.SR},
       adsurl = {https://ui.adsabs.harvard.edu/abs/2019arXiv190701877S}
}
\bibliographystyle{aasjournalv7}



\end{document}